\documentclass[fleqn, usenatbib]{mnras}

\usepackage{newtxtext,newtxmath}

\usepackage[T1]{fontenc}
\usepackage{ae,aecompl}

\usepackage{graphicx}	
\usepackage{amsmath}	
\usepackage{amssymb}	
\usepackage{lscape}
\usepackage{xcolor}
\usepackage{soul}

\usepackage{longtable,booktabs}

\title[AARTFAAC Flux Calibration and Catalogue at 60 MHz]{AARTFAAC Flux Density Calibration and Northern Hemisphere Catalogue at 60 MHz}

\author[M. J. Kuiack et al.]{
Mark Kuiack,$^{1}$\thanks{E-mail: m.j.kuiack@uva.nl}
Folkert Huizinga,$^{1}$
Gijs Molenaar,$^{3}$
Peeyush Prasad,$^{1}$\newauthor
Antonia Rowlinson,$^{1,2}$
Ralph A.M.J. Wijers$^{1}$
\\
$^{1}$ Anton Pannekoek Institute, University of Amsterdam, Science Park 904, 1098 XH Amsterdam, The Netherlands \\
$^{3}$ Department of Physics and Electronics, Rhodes University, PO Box 94, Grahamstown, 6140, South Africa \\
$^{2}$ ASTRON, The Netherlands Institute for Radio Astronomy, Postbus 2, 7990 AA, Dwingeloo, The Netherlands \\
}

\date{Accepted XXX. Received YYY; in original form ZZZ}

\pubyear{2018}

\begin{document}
\label{firstpage}
\pagerange{\pageref{firstpage}--\pageref{lastpage}}
\maketitle

\begin{abstract}

We present a method for calibrating the flux density scale for images generated by the Amsterdam ASTRON Radio Transient Facility And Analysis Centre (AARTFAAC). AARTFAAC produces a stream of all-sky images at a rate of one second in order to survey the Northern Hemisphere for short duration, low frequency transients, such as the prompt EM counterpart to gravitational wave events, magnetar flares, blazars, and other as of yet unobserved phenomena. Therefore, an independent  flux density scaling solution per image is calculated via bootstrapping, comparing the measured apparent brightness of sources in the field to a reference catalogue.  However, the lack of accurate flux density measurements of bright sources below 74 MHz necessitated the creation of the AARTFAAC source catalogue, at 60 MHz, which contains 167 sources across the Northern Hemisphere. Using this as a reference results in a sufficiently high number of detected sources in each image to calculate a stable and accurate flux scale per one second snapshot, in real-time.
\end{abstract}




\section{Introduction}

The Amsterdam ASTRON Radio Transient Facility And Analysis Center
(AARTFAAC) is an all-sky radio monitor, built as a parallel
computational back-end to LOFAR \citep[the Low-Frequency Array;][]{2013A&A...556A...2V}. 
 It operates primarily in LOFAR's low
band ($10-90$\,MHz) with an all-sky field of view, but can also
operate in the high band ($110-240$\,MHz) albeit only with an HBA tile
field of view ($30^\circ$ FWHM at 150\,MHz). It can be used to
monitor the radio sky on time scales upwards of one second, as often as
is practicable within LOFAR observing constraints and data storage
limitations.  Its core science goal is to search for rare, bright
transients at the lowest radio frequencies, which have proved to be
quite elusive
\citep[see, e.g.,][]{2014MNRAS.438..352B,2015JAI.....450004O,2016MNRAS.458.3506R,2016MNRAS.459.3161C},
but a few have been found in the otherwise poorly explored regime
accessible to AARTFAAC
\citep{2005Natur.434...50H,2016MNRAS.456.2321s,2017MNRAS.466.1944M}:
time scales of seconds to hours, and fluxes above several jansky in
the LOFAR low band ($10-90$\,MHz). In this regime, coherently
emitting objects will dominate, and thus any sources found will
represent fairly extreme or exotic physics; \citep[see, e.g.,][]{2015MNRAS.446.3687P}.

However, many other applications are possible, such as detecting very
high-redshift EoR signals \citep{2017ApJ...838...65P}, monitoring the
state of the ionosphere and phenomena in it
\citep{2015GeoRL..42.3707L,2015MNRAS.453.2731L}, and monitoring meteor
showers  \citep[e.g,][]{2014ApJ...788L..26O}.  And of course, many 
terrestrial phenomena such as RFI, air
planes, and satellites are detected that need to be distinguished from
more distant astrophysical signals and imaging artifacts before
science analysis can start.

At the moment, AARTFAAC all-sky monitoring is limited to times when
LOFAR is is LBA mode\footnote{An upgrade is planned and funded that will
allow simultaneous LBA and HBA observations}, but more importantly, due
to the fact that the search for interesting objects has not yet been
automated, and storage and offline search of the very large volumes of
raw data it generates is not practical. In this paper, we describe the next
step in achieving the goal of continuous all-sky monitoring with
AARTFAAC. Previously, we described the basic
properties of the AARTFAAC
system and its real-time calibration and technical commissioning
\citep{2014A&A...568A..48P},
as well as the system design and correlator \citep{2016JAI.....541008P},
and TraP, the transients detection pipeline also used for LOFAR transient
searches \citep{2015A&C....11...25S}.
In this paper, we describe the results of the first practical commissioning
runs. The basic goal of these runs was to collect enough data to cover
the full range of local sidereal times over a significant period of time
(so as to cover a range of ionospheric and RFI conditions) to investigate
practical strategies of bad data rejection, source extraction, and flux
calibration to enable future real-time operation. We collected over 30 hours of
data, which we will show is a good compromise between getting a manageable
dataset to experiment with extensively and sampling a sufficient range of 
conditions. In future work, we will first develop strategies for 
separating events of interest in large datasets from artifacts and known
variability, before implementing the full intended data analysis
(data taking, correlation, calibration, imaging, flux extraction,
feature recognition, and alert generation) into a streaming pipeline 
that can function in real-time.

To increase the sensitivity to fainter objects, the brightest
sources in the sky: Cygnus A, Cassiopeia A, Taurus A, Virgo A, The
Sun, hereafter referred to as the ``A-team", and most of the diffuse
Galactic plane emission, is removed during AARTFAAC calibration and
imaging. The images have, thus far, not yet been properly flux
calibrated \citep{2014A&A...568A..48P}. Instead the pixel values in
the resulting images are relative, with a scaling related to a
normalization of the total power received before ``A-team"
subtraction. An accurate characterization of transient phenomena
however, requires reference to a common physical scale. While studies of variability also require that each
extracted source measurement is made within a comparable reference
frame. This is only possible once each image has been corrected such
that the pixel values refer to the physical units of flux density,
janskys.

Radio flux density calibration is done by reference to catalogues of
stable, well studied calibrator sources. For example, a typical radio
observation includes observing a calibrator source before and after
observing the target. So the scaling of the calibrator data, the gain
solution, is applied to the target data.  In that case the gain
solution is assumed not to have changed substantially on the timescale
of the observation.

Unfortunately for AARTFAAC, the only large surveys below 100\,MHz are
The Very Large Array Low-frequency Sky Survey Redux
\citep[VLSSr;][]{2014MNRAS.440..327L} at 74\,MHz and The Eighth
Cambridge (8C) Survey \citep{1990MNRAS.244..233R} at 38\,MHz.  There is
therefore a gap across nearly the entire frequency range of AARTFAAC.
This clearly represents an opportunity for AARTFAAC to make an
important contribution, with unique flux density measurements of the
brightest sources in the $38-74$\,MHz range.

We therefore report on the method used to flux density calibrate
AARTFAAC images in real-time, for our future transient search
campaigns, and the resulting catalogue of bright sources.

Firstly, in Section \ref{sec:data} particular technical details of
AARTFAAC and the calibration observations are given.  Secondly, in
Section \ref{sec:calmethod} we describe the catalogue bootstrapping
method used to accurately flux density calibrate the images in
real-time.  
Then, we report on the
performance and stability of the method in section
\ref{sec:performance}, and analyses of the typical systematic
uncertainties.  Next in Section \ref{sec:cat} the characteristics  of
the first AARTFAAC catalogue of persistent sources at 60 MHz are
discussed.  And lastly conclusions are given in section
\ref{conclusion}.

\section{Data description}
\label{sec:data}

\begin{table*}
\centering
\begin{tabular}{l|l|l|l|l}
\hline \hline
 Start Date   &	Start - End  	& Start - End 	& Good Images & Un-flagged data\\
 &   	 [UTC]				&	[LST]	&	\#		&	\%					\\ \hline
2016 Aug 31 &   15:10	-   17:43	&	14h18m - 16h52m	&	8839	&	96.3	\\
2016 Sep 05 &   16:47	-   19:45	&	16h15m - 19h14m	&	10358	&	97.0	\\
2016 Sep 07 &   03:40	-   09:38	&	03h14m - 09h14m	&	21291 	&	99.1	\\
2016 Sep 30 &  	09:31	-   11:23	&	10h36m - 12h29m	&	2703	&	40.2	\\
2016 Nov 12 &  	06:32	-	19:53	&	10h26m - 23h50m	&	40145 	&	83.5	\\
2016 Nov 13 &  	20:00	-	22:57	&	00h01m - 02h58m	& 	5031	&	47.4	\\
2016 Nov 14 &	08:27	-	15:33	&	09h03m - 16h56m	&	23084	&	90.3	\\
2016 Dec 10 &	22:55	-	02:49	&	04h43m - 08h37m	&	9794	&	70.0	\\ \hline
Total &	32:56:46	&  &	121245	\\ \hline \hline
\end{tabular}
\caption{ The set of observations used to test the flux density calibration method and generate the first AARTFAAC catalogue at 60 MHz. The start and end of each observation are given as the UTC of the first and last image, as well as the LST centre of the image, both to the nearest minute.  During an observation data blocks may be flagged and removed either by the correlator, visibility calibration, or imaging pipeline. Then, the images were filtered based on the average image pixel RMS. Outliers are clearly the result of improper calibration, poor A-team subtraction, or bright RFI.}
\label{table:obs}
\end{table*}

By creating an all-sky image every second, AARTFAAC has the capability of generating a large amount of data. Therefore, the intended operational mode is to perform a transient search on the stream of images, saving only those data where an interesting event has been detected. However, in order to test the calibration method and fully characterize the data quality a set of observations was recorded and stored for analysis offline. 

Additionally, full LST coverage was required to generate the catalogue of calibrator sources across the Northern Hemisphere. Therefore, nearly 33 hours of observations were recorded to test the flux density calibration method and generate the AARTFAAC catalogue. This allowed the analysis of sources for many hours, across separate observations, while maintaining a manageable data volume. These observations were recorded between August and December of 2016, as outlined in Table \ref{table:obs}. During this period of time the final stages of commissioning with the real-time imaging pipeline were completed,  leaving only the image calibration.

In its present form the AARTFAAC system shares the 6 core stations, known as the ``Superterp," with  LOFAR, which is located near the village of Exloo in the Netherlands.  It was designed to operate in parallel with regular LOFAR observations by splitting the antenna signals from the stations and rerouting them to the AARTFAAC correlators and imaging servers. 

Each core LOFAR station consists of two sub arrays: the High Band Array (HBA), which has a bandpass from 120-240 MHz, and  the Low Band Array (LBA), with a bandpass from 10-90 MHz. AARTFAAC currently only uses the LBA.

These LBA  stations are  made of 96 pairs of orthogonal droop dipoles distributed with a roughly Gaussian density distribution. Their simple antenna design, two wires attached at 45 degrees to a central post over a metal mesh ground plane,  offer  a full sky field of view. Unfortunately, due to current computational constraints only data from 48 of the 96 antennas are processed. This subset of antennas may be distributed in one of 4 operating  modes:
\begin{itemize}
\item INNER: Antennas within 30 meters of the station centre.
\item OUTER: Antennas 30 to 87 meters.
\item 2 SPARSE modes: Either odd or even numbered antennas distributed throughout the station.
\end{itemize}

LOFAR LBA observes predominately in the OUTER configuration due to the larger number of longer baselines providing better UV filling of the superterp. In comparison using the INNER configuration results in dipoles which are more tightly clumped in the centre of the station, leaving more space between the stations. The OUTER configuration utilizes the outer ring of station dipoles which maximizes point source sensitivity and reduces diffuse background emission.  Additionally, regular LOFAR LBA observations will sum the antenna signals with a phase delay applied for the target pointing. However, because these phase delays are not applied at the stations during LBA observations, AARTFAAC has access to the raw signal from all 48 dipoles in operation, and is sensitive to the entire visible sky during all LBA observations.

The physical specifications of AARTFAAC are summarized in Table \ref{tab:specs}. Currently, in the standard operating mode of AARTFAAC a one second integrated Stokes I (1024x1024 resolution, 4.1 MB) fits image is created every second by integrating all 16 available subbands. This is a reduction from the initial total raw visibility rate 660 MB/s, including all subbands, which is reduced to 10 MB/s after calibration by averaging the 63 frequency channels which comprise each LOFAR subband. These calibrated visibilities are stored in the AARTFAAC archive for offline processing, and the upcoming transient survey. However, in the future only those data which are found to contain an interesting transient event will be stored.

In order to maximize sensitivity while reducing RFI, the subbands are configured in two continuous sets of eight subbands, 57.52 - 59.08 MHz and 61.04 - 62.6 MHz. This is near enough the peak sensitivity around 57 MHz \citep{2013A&A...556A...2V}, while avoiding frequencies which have been observed to have a higher RFI occupancy percentage,  \citep[see Fig. 6,][]{2013A&A...549A..11O}. With this configuration a pixel RMS < 10 Jy is achieved over 40\% of the Northern Hemisphere, while 90\% achieves RMS < 21 Jy.

\begin{table*}
\centering
\begin{tabular}{l|l|l}
\hline \hline
 Parameter   			&	AARTFAAC LBA   			& Comment \\ \hline
Array elements 			&   288  inverted V antennas &	Dual polarized elements		\\ 
Frequency range  		&   10-90		(MHz)			&		\\
Field of view 			&   $\pi$		(sr) 			&	FWHM of beam	\\
Total Effective area 	&   $2617^a$	($m^{2}$)			&		\\
$T_{\mathrm{sys}}$  		&  	3600	($\nu^{-2.55} $K)			&		\\
Angular resolution 		&  	60 	(arcmin) 				&		\\
Subband resolution 		&  	195  	(kHz) 					&		\\
Processed Bandwidth  	&	3.12	(MHz)					&		\\
Temporal resolution 	&	1	(s) 					&		\\ \hline \hline
\end{tabular}
\caption{ AARTFAAC system design specifications, from Table 1 of \citep{2014A&A...568A..48P}. Here the subband and processed bandwidth values are updated to reflect the current operational capabilities of AARTFAAC.}
\label{tab:specs}
\end{table*}

\section{Image Calibration}
\label{sec:calmethod}

The AARTFAAC real-time calibration and imaging pipelines, as they are currently implemented, output all sky snapshot images at a rate, and integration time, of one second \citep{2014A&A...568A..48P}. Yet, before the images can be used for transient detection or variability analysis, two corrections must be made: First, a direction dependent rescaling, which corrects the images based on the antenna response pattern, also known as the primary beam. Then, a direction independent rescaling, which transforms the pixel values from arbitrary units to a flux density in janskys.

Both of these corrections are important before the image stream can be analyzed for transient or variable sources. Clearly, without accounting for any variation in the antenna response across the sky, where sensitivity peaks at zenith, the brightness of all sources would appear variable as they track across the field of view. 

Similarly, applying a reliable flux density scaling to each image is vital for measuring variability. Given that the pixel values in raw images are arbitrary, with an unknown influence from the subtraction of the A-team sources, it would be difficult to determine whether any variation is intrinsic or an artifact of the calibration and imaging. Furthermore, searching for a transient source in an image with arbitrary scaling would make it impossible to determine the shape of the light curve, spectral index, or whether a candidate is indeed astrophysical at all. 

These characteristics are also critical when devising further observations as they dictate what sensitivity and spectral coverage are necessary to maximize the likelihood of a follow-up detection. Additionally, in the cases where detections are made but follow-up observations yield no result, the  shape and flux density distributions of transient light curves would be useful to model possible progenitor populations, and to compare our results with those of other low frequency surveys. This is the case for many FRB studies to date \citep{2015MNRAS.447..246P}. 

\subsection{Beam Model}
\begin{figure*}
\begin{center}
\includegraphics[width=\textwidth]{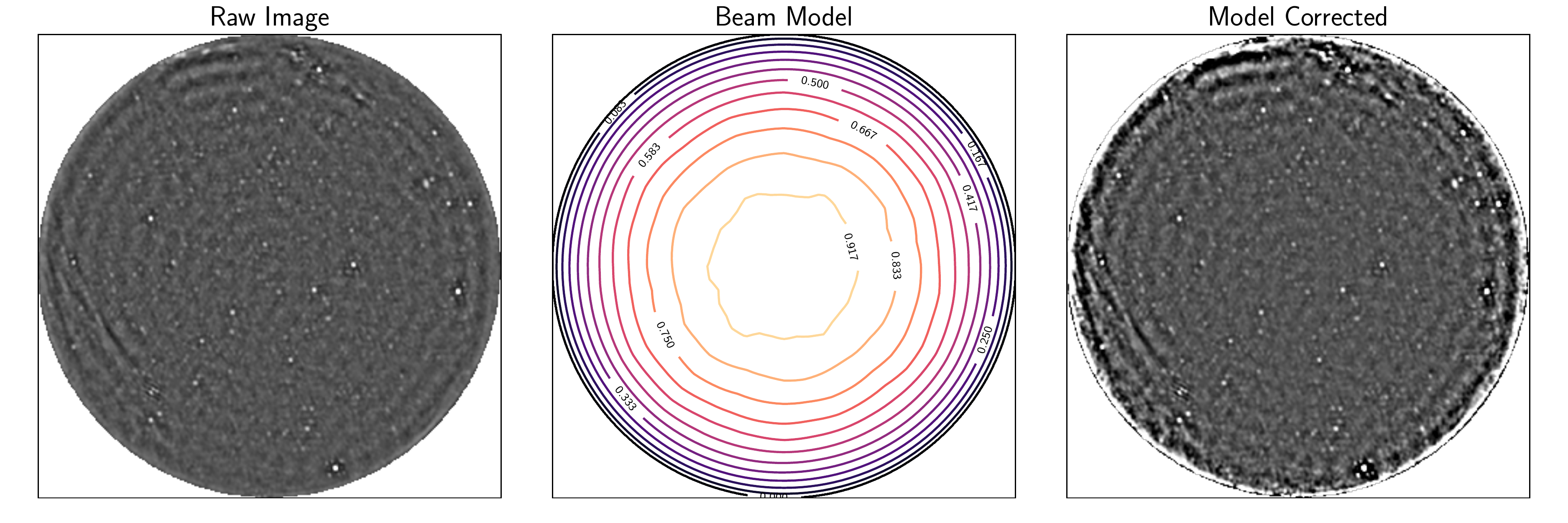}
\caption{Left: An AARTFAAC all-sky image, output by the imaging pipeline, after A-team subtraction. Centre: A normalized beam model, for AARTFAAC at 60MHz. The model shows the shape of the direction dependent gain across the field of view with maximum gain at zenith, decreasing toward the horizon. Right: the same image with the correction applied. Note, these images have not be flux density calibrated so  the pixel scaling is arbitrary. } 
\label{fig:beamsim}
\end{center}
\end{figure*}

The beam model is an approximation of the direction dependent sensitivity, across the field of view, of the array. Thus, correcting for this differential gain response pattern ensures that the light curves of detected sources are flat, as they move across the field of view. 

This pattern is the result of many physical factors including: observing frequency, the geometry of the stations and dipoles, their mutual interactions with each other, and the effect of the local terrain. 
Given these complicated interactions, it is modelled by simulating the full station layout of all of the dipoles which form the six stations on the LOFAR Superterp, with accurate placement and orientation, across the frequency spectrum.

\begin{figure}
\begin{center}
\includegraphics[width=0.5\textwidth]{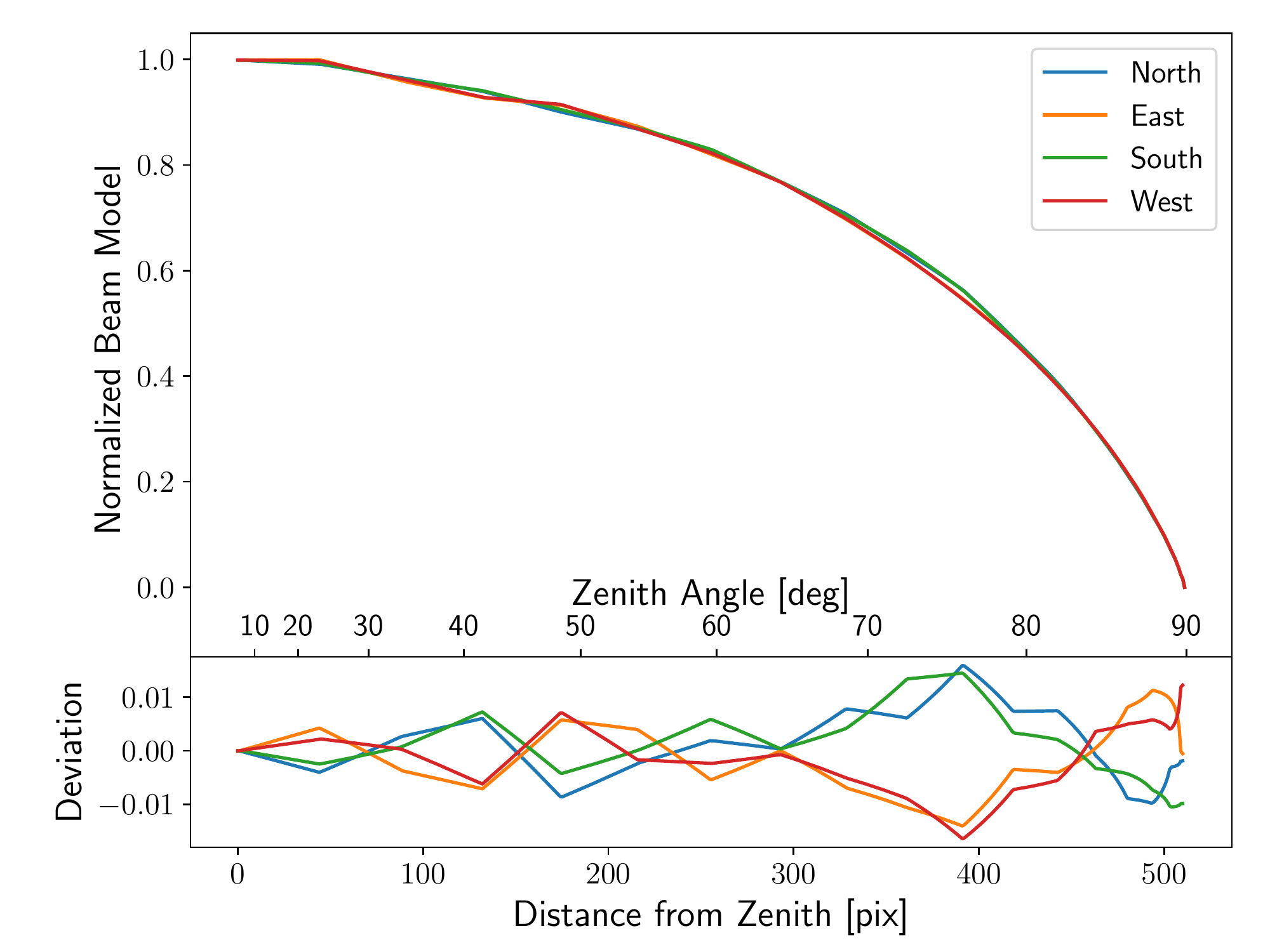}
\caption{Comparing the beam model gain pattern at 60 MHz, from zenith to horizon, in the four compass directions. The values are normalized to their maximum value, which is at zenith. The beam is highly symmetric about the zenith, with relative deviations from perfect symmetry below 1\% out to $70^{\circ}$.  } 
\label{fig:beamshape}
\end{center}
\end{figure}

Therefore, the beam response shape has been modelled at frequencies between 30 and 70 MHz in 5 MHz intervals. This covers the spectral range of AARTFAAC with sufficient accuracy, since the model does not change rapidly with frequency. These models were generated using WIPL-D, an electromagnetic simulation software package. Additionally, Fig. \ref{fig:beamshape} illustrates the symmetry of the model at 60 MHz, about the zenith.

\begin{figure}
\begin{center}
\includegraphics[width=0.5\textwidth]{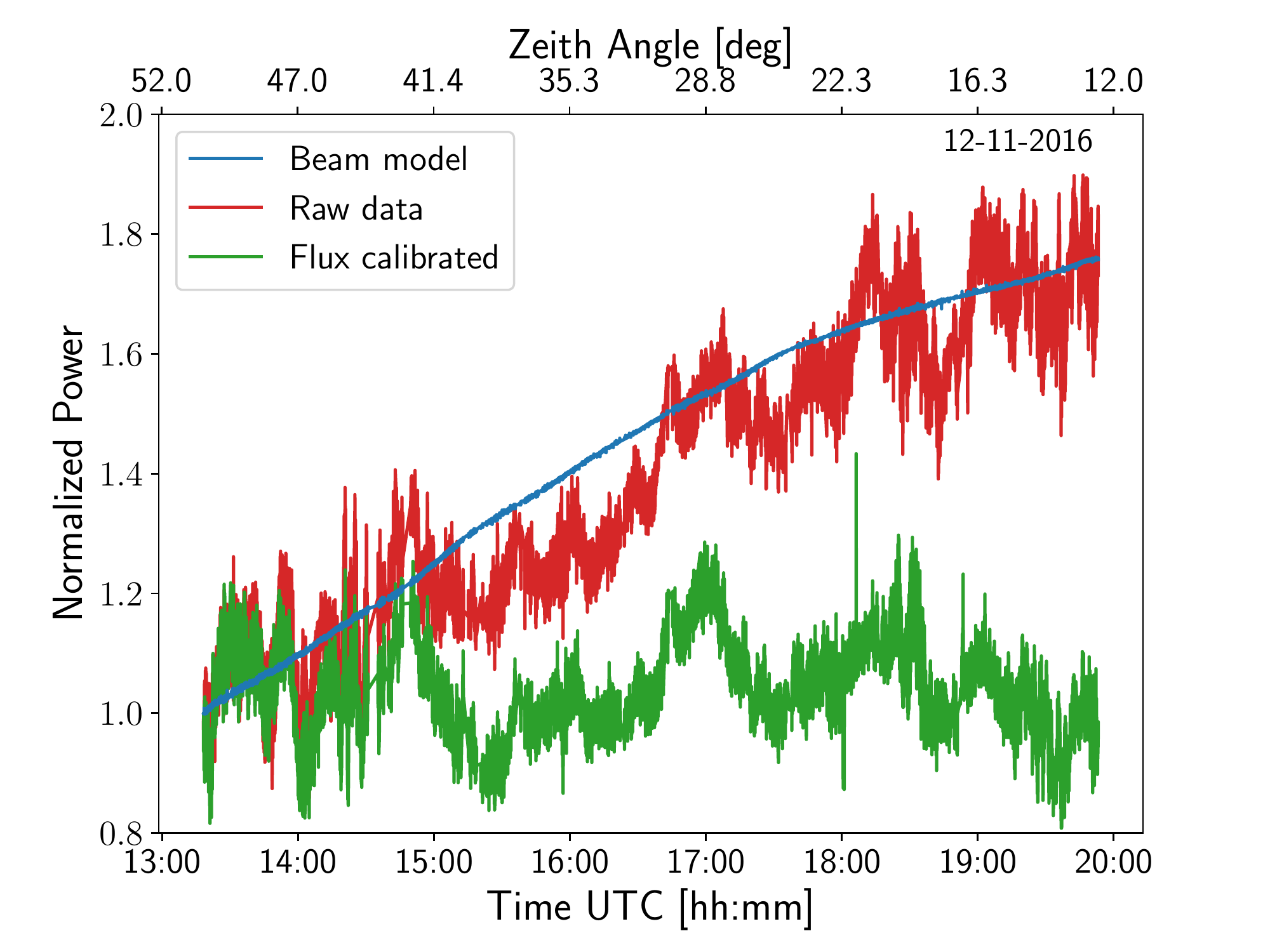}
\caption{Example light curve for a source observed for $\sim 7$ hours, normalized to the first data point. The raw data (red) before calibration clearly shows the shape of beam pattern along the track of the source across the sky (blue), while after calibration (green) the curve is much flatter.  } 
\label{fig:beamexample}
\end{center}
\end{figure}

The left image of Fig. \ref{fig:beamsim} shows an example raw AARTFAAC image. Although the background and noise appear flat across the image, sources decrease in apparent brightness as the sensitivity drops toward the horizons. 
The sensitivity peaks at zenith and decreases toward the horizon. Therefore the images are corrected by dividing the raw image by the image of the beam model, normalized such that the gain at zenith is 1. The shape of the normalized beam model at 60 MHz is illustrated in the middle panel of Fig. \ref{fig:beamsim}, and the resulting beam corrected image is shown on the right.
Given that the sensitivity near the horizon is lower, after the correction is applied, the noise near the horizon is also multiplied. However, the mean flux density of sources will be constant as they rotate though the beam, as illustrated in Fig. \ref{fig:beamexample}.

Although the beam model has been observationally verified during LOFAR commissioning \citep{2013A&A...556A...2V}, AARTFAAC is able to perform an additional test, using sources detected across the field of view, observed over hours as they move across the sky.  After calibrating the data, the extracting flux measurements of each source at different locations on the sky were compared to the mean of their light curve. No position dependent deviations, which would indicate an improperly modelled beam, were observed within our detection region of 50 degrees from zenith.

\subsection{Reference Catalogues}
\label{sec:refcat}

Given the real-time streaming nature of AARTFAAC observations, an algorithm that computes the flux scaling per image, using only the image itself, is preferred. This excludes traditional flux density scale calibration techniques, such as intermittently observing another field with a calibrator source. 

Fortunately, with our field of view encompassing the entire sky, enough bright sources are present in each image to compute the scaling in real-time.  However, this is only possible if accurate apparent fluxes at the observing frequency are known. Therefore, with a population of enough bright sources covering the Northern Hemisphere, the apparent brightness of those sources extracted in each image can be compared with the expected flux density, and the conversion scale factor computed. This is described fully in Section \ref{sec:fluxmethod}.

Furthermore, utilizing the greatest number of calibrators ensures that the variability of any one source does not dominate the calibration solution. Additionally, ionospheric fluctuations are phenomena localized on the sky, as density fluctuations travel through the field of view, and therefore decorrelate on larger angular sizes.  To this end, gathering together a catalogue of all sources with a signal-to-noise ratio > $5\sigma$, and broadband spectra across our entire observing range, and field of view, would allow accurate and stable flux density calibration at any observing frequency.

Recently, several catalogues of calibrator sources have been published with accurate broadband spectra in the LOFAR LBA range, 30-80 MHz: 

One example, \cite{2012MNRAS.423L..30S} contains six bright sources from the Third Cambridge Catalogue, 3C, with spectral models between 30 and 300 MHz. Unfortunately, with only six sources spanning the Northern Hemisphere, AARTFAAC images would not contain enough calibrator sources to ensure a robust scaling solution at all observing times. However, the analytic spectral models across the full LBA band and the fact that these sources appear in the other catalogues adds a useful inter-catalogue check. 

Secondly, \cite{2017ApJS..230....7P} published modelled spectra for 20 sources with flux density measurements down to 74 MHz. Four of these are the A-team sources which are subtracted from the images before this flux density calibration step, leaving 16 sources. Of these, 6 have a declination $< -10^{\circ}$, and are therefore outside, or too near the edge, of our field of view. This leaves 8 sources which are viable flux density calibrators. Even still, rarely are more than 3 of these sources visible in the sky simultaneously,  which is preferred to have stable flux density fit solution at all times.

Lastly, the catalogue published by \cite{2008ApJS..174..313H} contains spectra for 388 sources selected from VLSSr which are brighter than 15 Jy at 74 MHz. The spectral shapes of all sources are described with either a single power-law, or if enough data are present, by the function $Y = A + BX + C \exp(DX)$, where $Y = \log ( F_{\nu} / 1 ~\mathrm{Jy})$ and $X = \log (\nu / 74 ~\mathrm{MHz})$,  which describes a spectral turnover of the flux density, $F_{\nu}$, at lower frequencies, $\nu$, in some sources. 

In addition to these, the full VLSSr catalogue \citep{2014MNRAS.440..327L} was used to follow up sources which are  detected in AARTFAAC images with $ >5\sigma$ signal-to-noise but are not associated with any object in the \cite{2008ApJS..174..313H} catalogue. These might result from two or more sources in VLSSr with < 15 Jy that are sufficiently close together to be confused at AARTFAAC's resolution. 

The lower resolution of AARTFAAC and the densely packed array allows us to see much more diffuse emission than the VLSSr. For this reason, the supernova remnant catalogue by \cite{2014BASI...42...47G} was also used for source association. These objects are of interest to us because they are bright at low frequencies, and their larger angular size reduces the effect of ionospheric scintillation. However, due to the frequency at which these flux densities are given (1.4GHz) and the much narrower beam width, it is impossible to simply extrapolate and compare with AARTFAAC measurements. 

The spectral models published by \cite{2017ApJS..230....7P} are much more accurate in the frequency range observed with AARTFAAC. However, this catalogue does not contain enough sources across the Northern Hemisphere to ensure that 3 or more sources are observable simultaneously, which is a requirement, used to ensure a more stable scaling solution in the presence of scintillation. And while \cite{2008ApJS..174..313H} present a catalogue with many more sources, the simpler spectral models result in a much greater uncertainty in the flux density predictions below 74 MHz. Neither catalogue was therefore sufficient to accurately compute a flux density scale at all times. This necessitated the creation of the AARTFAAC low frequency catalogue, which is outlined in detail in Section \ref{sec:cat}.

\subsection{Flux density scale}
\label{sec:fluxmethod}

\begin{figure}
\begin{center}
\includegraphics[width=0.5\textwidth]{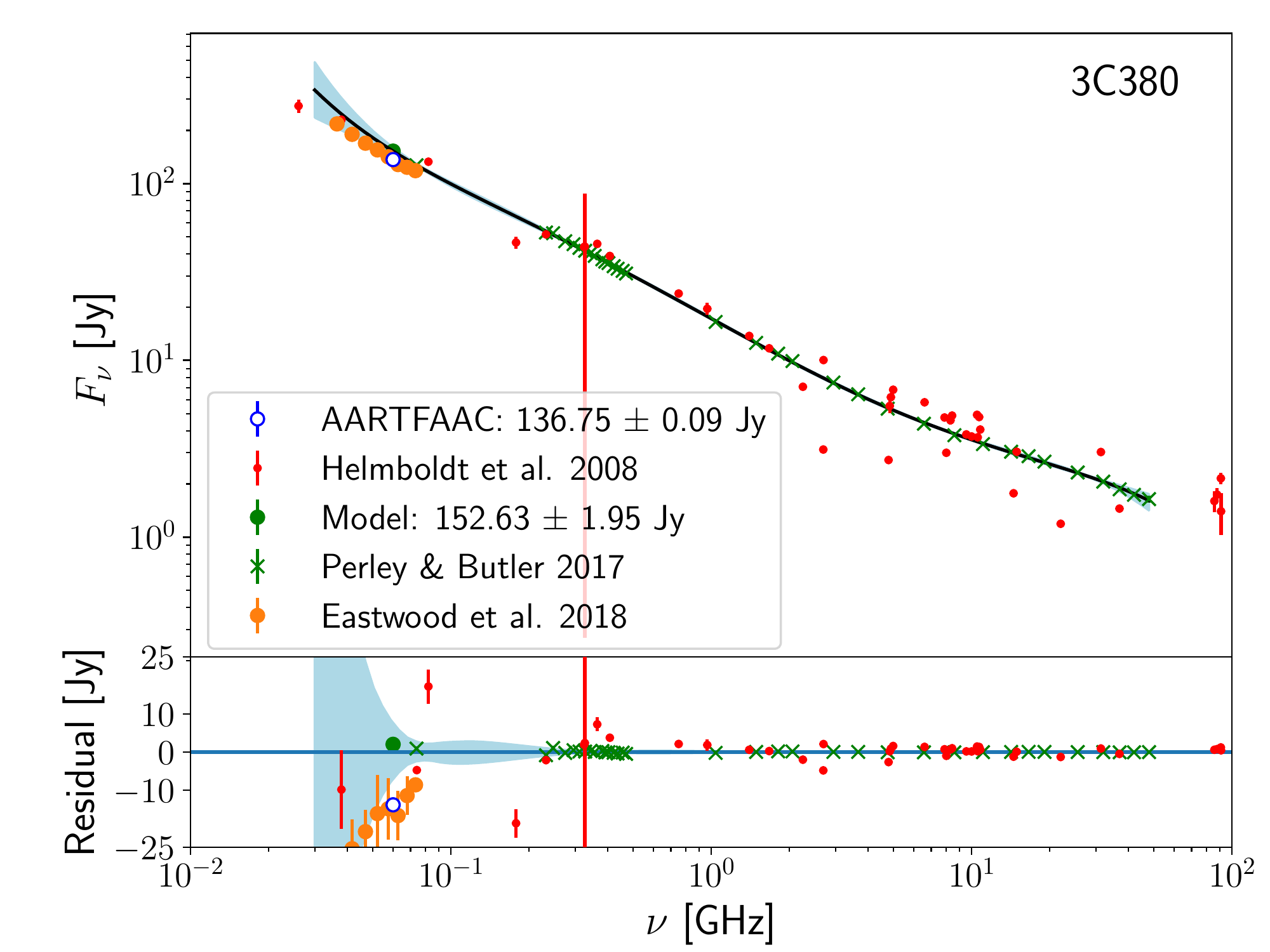}
\caption{ The spectral model of 3C380 is generated from the flux density measurements by \citet{2017ApJS..230....7P}. The uncertainty in the model (light blue region) is calculated via Monte Carlo random sampling of sets of measurements based on their reported uncertainty. This gives the model value at our measurement frequency, 60 MHz (green filled circle). The AARTFAAC catalogue flux density value (open circle) and \citet{2008ApJS..174..313H} measurements (red dots) and \citet{2018AJ....156...32E} measurements (orange filled circles) are compared to the model (black line). The residuals show the difference between the measured values and the model.
}
\label{fig:3c380_spectra}
\end{center}
\end{figure}

The AARTFAAC catalogue was created by first modeling the \cite{2017ApJS..230....7P} source fluxes at lower frequencies. This was done by using a Monte Carlo method where 10,000 sets of flux density measurements for a given source at different frequencies were generated using the flux density measurements and uncertainties available in the supplementary online data. Next a least squares fit was done to the spectral models with the same polynomial degrees as those in \mbox{\cite{2017ApJS..230....7P}}: 
\begin{equation}
\log \mathbf{(F_{\nu})} = a_{0} + a_{1} \log (\nu_{G}) + a_{2} \log (\nu_{G})^2 + a_{3} \log (\nu_{G})^{3} + ...
\end{equation}
where $F_{\nu}$ is the flux density in Jy and $\nu_{G}$ is the frequency in GHz. Finally, the resulting functions are then evaluated at the AARTFAAC observing frequency. This provides predicted fluxes with accurate uncertainties. The spectral model for one such source, 3C380, is shown in Fig. \ref{fig:3c380_spectra}, along with the flux density measurements from \cite{2008ApJS..174..313H} and \cite{2017ApJS..230....7P}. Here, the model predicted flux density at 60 MHz, and the final AARTFAAC catalogue flux density,  after averaging the data from all observations, is shown not to be in agreement, within their mutual uncertainties. However, when comparing with the flux density as measured by \citet{2018AJ....156...32E}, the differences are much smaller for all sources, except 3C286.  Therefore, because of the similarity of the LOFAR-LBA and OVRO-LWA antenna design, but different imaging and calibration method,  it is clear that the measured flux density at 60 MHz is accurate.

The AARTFAAC catalogue is then made by bootstrapping the flux density scale from these sources in the following way: 

First, an observation of a few hours was considered at a time when 5 calibrator sources were visible. 

Then the predicted flux density values at 60 MHz, obtained from the reference catalogue, were compared to their sigma-clipped light curves extracted from the observation. Iteratively clipping the flux density measurements greater than 3$\sigma$ from the mean reduces the possible effect of RFI or imaging artifacts. 

Light curves are generated by extracting the source fluxes from each image with the Python Source Extractor \citep{2018A&C....23...92C}, then the sources extracted from each image are associated with the extractions from previous images in a running catalogue database. This is performed using the LOFAR Transient Pipeline \citep[TraP;][and references therein]{2015A&C....11...25S}.

The durations of the observations used for generating the catalogue were each longer than 2 hours. This ensures that scintillation effects shown in Fig. \ref{fig:beamexample}, which occur on a 15-20 minute time scale \citep{2015RaSc...50..574L}, are averaged out. 

We then calculate the scaling solution via linear regression, weighting the sources according to their measurement uncertainty. The single scale factor calculated for that observation was then applied to the mode of the flux density measurements of the other persistent sources, detected above $5\sigma$. 

Those inferred values and our measurement of the original reference sources are added to the AARTFAAC catalogue. By adding the additional calibrators, and using the new AARTFAAC catalogue as a reference for the other observations where fewer \cite{2017ApJS..230....7P} calibrators are visible, a more accurate scale factor for each additional observation can be calculated. 

To summarize, in each new observation, all light curves longer than 2 hours are extracted, the scale factor for the observation is calculated, and it is applied to the new sources, then the new sources are added to the catalogue. The number of data points for each source is recorded so that when a source, which already exists in the catalogue, is re-observed, a mean weighted by the number of measurements is used to calculate the updated flux density value.

Now that the entire Northern Hemisphere has been observed, and all persistent sources detectable above 5$\sigma$, flux densities of 40-50 Jy for much of the sky, have been added, the AARTFAAC catalogue can be used to flux density calibrate any individual AARTFAAC image. This is possible because there are are 30-50 observable sources above this threshold at any time. This greater number of sources, across the full field of view, provides a stable flux density solution.

Finally, the AARTFAAC catalogue was used to correct the flux density scale on the full 33 hours of data. This was done in a streaming mode, calculating the the linear scaling solution for each image independently, so that the intended use case could be analyzed by verifying light curves extracted from the data. In doing so we observed that the scaling solution did have a dependence on LST, which was expected due to the scaling of the raw visibilities according to the total power of objects in the field of view, but that the scaling solutions, at a given LST, were stable across the months spanned by the set of observations.

\section{Analysis of Flux calibrated data}
\label{sec:performance}

AARTFAAC produces a snapshot image of the entire sky at a rate of 1 per second. To this stream of images a correction for the antenna response pattern, as well as a scale factor, per image, is applied. This enables the creation of a reliable light curve for each source. In the previous section, the method for correcting each image by applying the beam model and calculating the scale factor, required to scale the pixel values to flux densities in janskys, was discussed.

Each of these corrections influence the light curves on different time scales: the varying sensitivity of the antenna will modulate the apparent brightness over a long period of time as the source moves across the sky, whereas the flux density scaling is corrected on each image independently and would therefore have its effect on the timescale of ionospheric fluctuations.

\begin{figure*}
\begin{center}
\includegraphics[width=\textwidth]{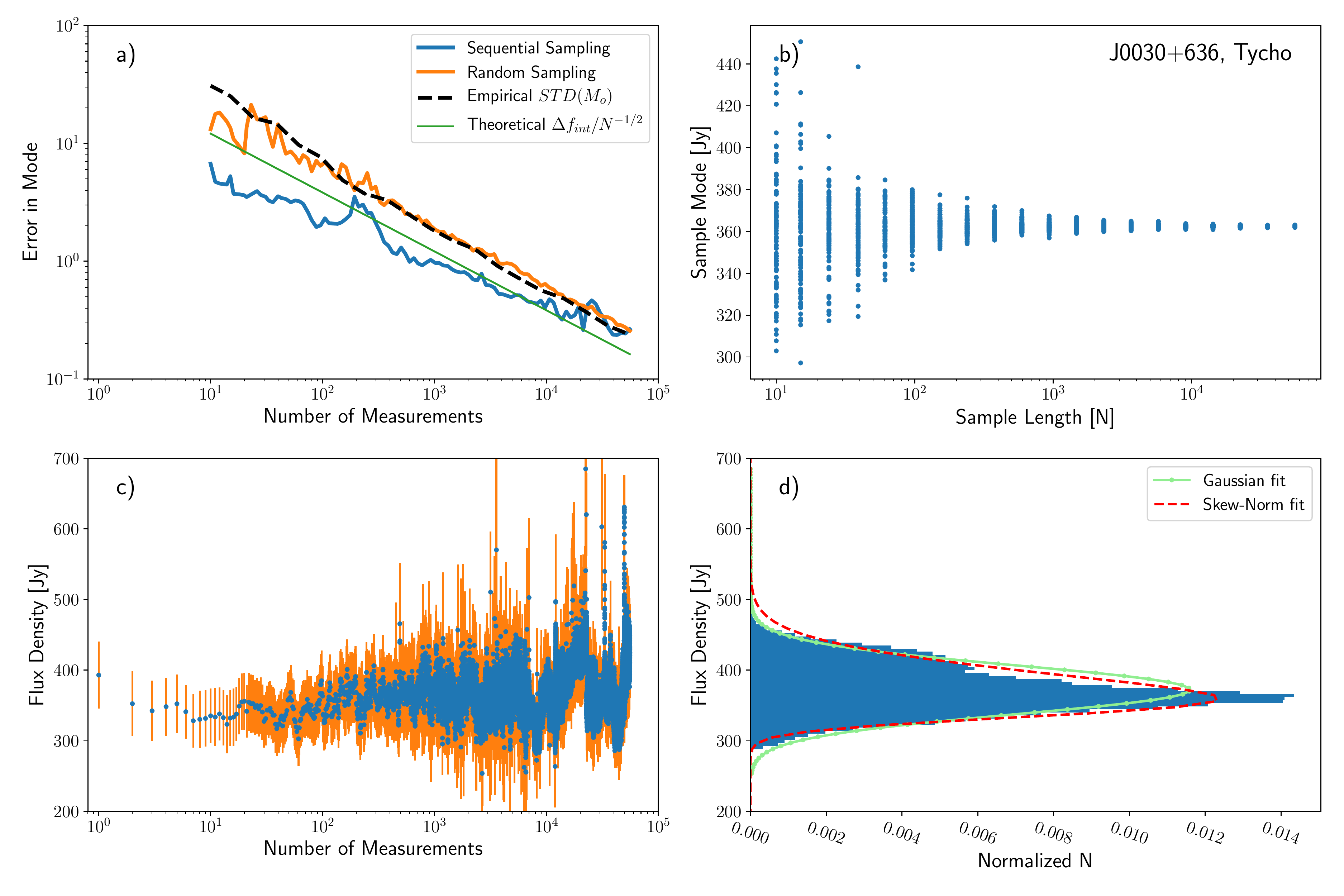}
\caption{ The noise profile of the Tycho supernova remnant. The source is intrinsically stable due to its large size, and emission mechanism, so fluctuations in its measured flux density are due strictly to either source fit statistics or ionospheric fluctuations. Figure a shows the error in the mode as a function of sample size, for different methods of sampling the data, and different methods of estimating the uncertainty. The sequential sampling curve (blue) was generated  by taking subsets of the measurements from the light curve in order. This shows how the uncertainty is affected by the systematic sources of error over time. The random sampling curve (orange) was generated by randomly sampling, with replacement, from the entire light curve and estimating the uncertainty in the mode of that subsample. The empirical (black dashed) was calculated by randomly sampling multiple subsets of the light curve, then calculating the standard deviation in the modes of those subsets, then repeating this for a number of sample lengths. These populations of mode, from different subsets, are plotted in figure b.  Lastly the theoretical line (green) shows how the standard error in the mean value of the data would be expected if the noise were purely Gaussian, independent and equal to the average uncertainty in the fit from each image. Figure c shows the entire light curve of Tycho from our set of observations, with the uncertainty of each measurement in orange. Lastly, figure d presents a histogram of the data, and a comparison to a normal and skew normal distribution.
} 
\label{fig:tycho_noise}
\end{center}
\end{figure*}

Turbulence in the ionosphere causes the apparent brightness, as well as the position and shape, of sources to fluctuate. This reduces the instantaneous accuracy of measurements from individual AARTFAAC images. Fortunately, this is overcome by observing each source for a sufficiently long time that the mean value of the light curve converges. 

The length of time for which each source must be observed depends in general on the typical timescale of ionospheric scintillation. For example, if a source is observed many times, but for a shorter duration than the timescale of these fluctuations, the variance of the light curves, and thus the uncertainty in the flux density will be dominated by the ionospheric fluctuations. 

As an example Fig. \ref{fig:tycho_noise} illustrates this for the Tycho supernova remnant. Large sources of synchrotron emission, such as supernova remnants generally do not intrinsically vary in brightness, making them useful tools to study the systematic effects on our flux density measurements. 

For a pure Gaussian noise process the standard error in the mean, SEM, defined as the standard deviation of the means calculated from subsets of the data, decreases proportional to the number of samples in the subset, $\sigma_{m,N} \propto N^{-1/2}$. This is the green ``Theoretical'' line in Fig. \ref{fig:tycho_noise}a, scaled by the average uncertainty in each individual integrated flux density measurement, $\Delta f_{\mathrm{int}}$.

However, Fig. \ref{fig:tycho_noise}d illustrates that AARTFAAC flux density measurements are not a pure Gaussian process. In fact, despite the larger angular size of Tycho, its light curve, shown in Fig. \ref{fig:tycho_noise}c, reveals that the measured flux density is still significantly modulated by the ionosphere. 

There are therefore two regimes, timescales much less than, or much greater than the timescale of ionospheric fluctuations, $10^{2}$ - $10^{3}$ seconds, which represent the dominant sources of uncertainty in AARTFAAC flux density measurements. 

First, the statistical uncertainty in each individual source fit due to the image noise. For timescales much less than the fluctuations in the ionosphere (<$10^{2}$ seconds), these measurements are highly correlated and thus not independent. This is due to the fact that AARTFAAC images are confusion noise limited. The noise profile is therefore below what is estimated by the individual source fits. This is evident in the left side of Fig. \ref{fig:tycho_noise}a, where the ``Sequential sample'' curve, generated by calculating the mode of subsamples from the light curve sequentially over increasing time, is below the pure Gaussian estimate.

Secondly, the variation of the brightness due to the electron density fluctuations in the ionosphere. Again, the noise profile differs from the Gaussian estimate for light curves much longer than the typical ionospheric timescale ($> \sim 10^{2}$ seconds). The variation caused by the ionosphere causes fluctuations which are greater than the estimated uncertainties from each image. This effect causes the estimated error to cross above the Gaussian estimate.

Furthermore, observing a source for many hours will result in significant motion across the sky, and through the beam of the antenna. Fig. \ref{fig:beamexample} illustrates an example of correcting for the antenna response on a source, whose light curves have been normalized to the first data point. Clearly, in the raw light curve (red) the increasing sensitivity of the antenna is visible as the source rises in the sky toward zenith. Along side this, the beam response pattern along the path of the source (blue) illustrates how as the sensitivity increase toward zenith and the beam centre, explaining the dramatic increase in the apparent brightness.  When the beam model is divided out, a much flatter calibrated light curve (green) remains. In fact, the residual variability in the light curve shown is predominantly due to turbulence in the ionosphere, causing the apparent brightness fluctuations.

Consequently, these noise characteristics indicate that observations shorter than a few minutes duration, may not yield an accurate average flux density value, regardless of the noise properties in each individual image.
However, the histogram of flux density values measured from the entire observation, shown in Fig. \ref{fig:tycho_noise}d, illustrates that the noise profile is, by appearance, nearly Gaussian, after observing the source for a period significantly longer than the timescale of the ionosphere. This is to be expected given the central limit theorem. Therefore,  making prolonged observations results in both accurate and precise flux density measurements.

\subsection{Flux density measurement statistics}
\label{sec:cat_flux}

As the noise was expected to be Gaussian, with potential systematic influences from either the ionosphere or an incorrectly modelled beam pattern, a skewed normal Gaussian distribution was fit to each light curve.
\begin{equation}
\label{eqn:fullpdf}
f(x) = \frac{2}{\omega} \phi \left( \frac{x - \xi}{\omega} \right) \Phi \left( \alpha \left( \frac{x - \xi}{\omega} \right) \right), 
\end{equation}
where $\phi$ is the standard normal probability distribution and $\Phi$ is its cumulative distribution, and the transformation \mbox{$x \rightarrow \frac{x - \xi  }{\omega}$}, to the fitted parameters $\xi$, the location, $\omega$ the scale, and $\alpha$, the shape, which defines skewness. The skewness increases with the absolute value of $\alpha$, with more weight in the left tail when $\alpha < 0$ and in the right tail when $\alpha > 0$,  when $\alpha = 0$ the skewed normal distribution becomes the normal distribution, and $\xi$ is simply the mean, and $\omega$ the standard deviation.

After fitting Eqn. \ref{eqn:fullpdf} to each of the light curves, it was found that the skewness was most frequently positive, with a larger tail in the distribution towards higher flux. This could indicate that the variation, introduced by the ionosphere, preferentially focuses the light into shorter bright caustics. These move along the ground, similar to the light on the bottom of a swimming pool.

Consequently, simply integrating over long observations, either by simple average, or calculating the median would yield results biased by the preference for outliers with increased brightness. Hence, the mode of the distribution is the most robust measurement of the central tendency of each source, and therefore the most accurate description of its flux density. 

However, the mode of the skew normal distribution is not analytic, but can be approximated numerically,
\begin{equation}
\label{eqn:mode}
M_o \approx \xi + \omega m_o(\alpha),
\end{equation}
where the $\xi$, $\omega$, and $\alpha$, are the fit parameters location, scale, and shape, of the distribution. The function $m_o(\alpha)$ is the degree to which the skew modifies the mode from the mean, which for a normal distribution is equal to 0. This is given by, 
\begin{equation}
m_o(\alpha) \approx \mu_z - \frac{\gamma_1 \sigma_z}{2} - \frac{\mathrm{sgn}(\alpha)}{2} \exp \left( -\frac{2 \pi}{| \alpha |} \right).
\end{equation}
where   $\sigma_z = \sqrt{1 - \mu_z^2}$, such that 
\begin{equation*}
\mu_z = \sqrt{\frac{2}{\pi}} \delta,
\end{equation*} 
for 
\begin{equation*}
\delta = \frac{\alpha}{\sqrt{1+\alpha^2}},
\end{equation*}
and where $\gamma_1$ is the skewness, 
\begin{equation}
\gamma_1 = \frac{4 - \pi}{2} \frac{\left(\delta \sqrt{2/\pi} \right)^3}{ \left(1 - 2 \delta ^2 / \pi \right)^{3/2} }
\end{equation}

The parameters were fit using the Bayesian inference MCMC package PyMC3. This method randomly samples the parameters from defined prior distributions, then computes the likelihood.   
 The uncertainty estimate was output by the PyMC3 package \citep{2015arXiv150708050S}. By defining the mode as a deterministic function of the fit parameters, PyMC3 gives the resulting uncertainty of the mode directly, as well as producing a covariance matrix for the parameters, and  plots of the marginal posterior probability distributions.

The robustness of the mode and correctness of the uncertainty calculation was tested in two ways. 

First, subsets of the data of varying length were randomly sampled with replacement, from the light curve. Then the mode of each subset was calculated. Fig. \ref{fig:tycho_noise}b shows the variance in the mode of each sample of Tycho flux density measurements, taken from different length intervals of the total observation.  As the length of these intervals was increased, the standard deviation of the mode of each subset was calculated, and plotted as the dashed black line in Fig. \ref{fig:tycho_noise}a.  

Then, the ``Random Sampling'' curve in Fig. \ref{fig:tycho_noise}a shows the estimation of the uncertainty in the mode of a randomly drawn sample from the light curve. 
Comparing these uncertainty estimates, the PyMC3 estimate from the random sample, with the empirically measured standard deviation of the modes from a number of different subsets, and the estimate calculated from the fit parameter uncertainties added in quadrature, provides an additional independent verification of the reported uncertainty.

As previously argued this catalogue presents the mode of the skew normal distribution as the most robust measurement of the flux density of each source, under the influence of a turbulent ionosphere. The observed tendency toward positive skewness indicates that longer integrations, or simply averaging successive shorter integrations,  would yield a positive bias in the inferred flux density.

Given that AARTFAAC images are generated and calibrated at a rate of one per second, well below the typical time scale of ionospheric scintillation, it is possible to observe a large number of flux density measurements, fit the distribution, and calculate the mode. 
However, other low frequency surveys typically make much longer integrations to reduce the image noise level. Therefore the quantity which should be compared is the mean, rather than the mode. The mean, $\mu$, of the distributions fit can be calculated from the parameters given in the catalogue as follows, 
\begin{equation}
\mu = \xi + \omega \delta \sqrt{\frac{2}{\pi}}.
\end{equation}

Lastly, the uncertainty presented in the flux densities are the statistical uncertainty in the modes of each distribution. As shown in Fig. \ref{fig:tycho_noise}a follow up measurements would agree within the stated uncertainty if the duration of the observations is sufficient. However, observations shorter than the ionospheric time scale could only be expected to agree within the standard deviation, described by,  
\begin{equation}
\sigma = \sqrt{\omega^2 \left(1- \frac{2 \delta^2}{\pi} \right)  }.
\end{equation}

\section{Catalogue }
\label{sec:cat}

\subsection{General properties}

\begin{figure*}
\begin{center}
	\includegraphics[width=\textwidth]{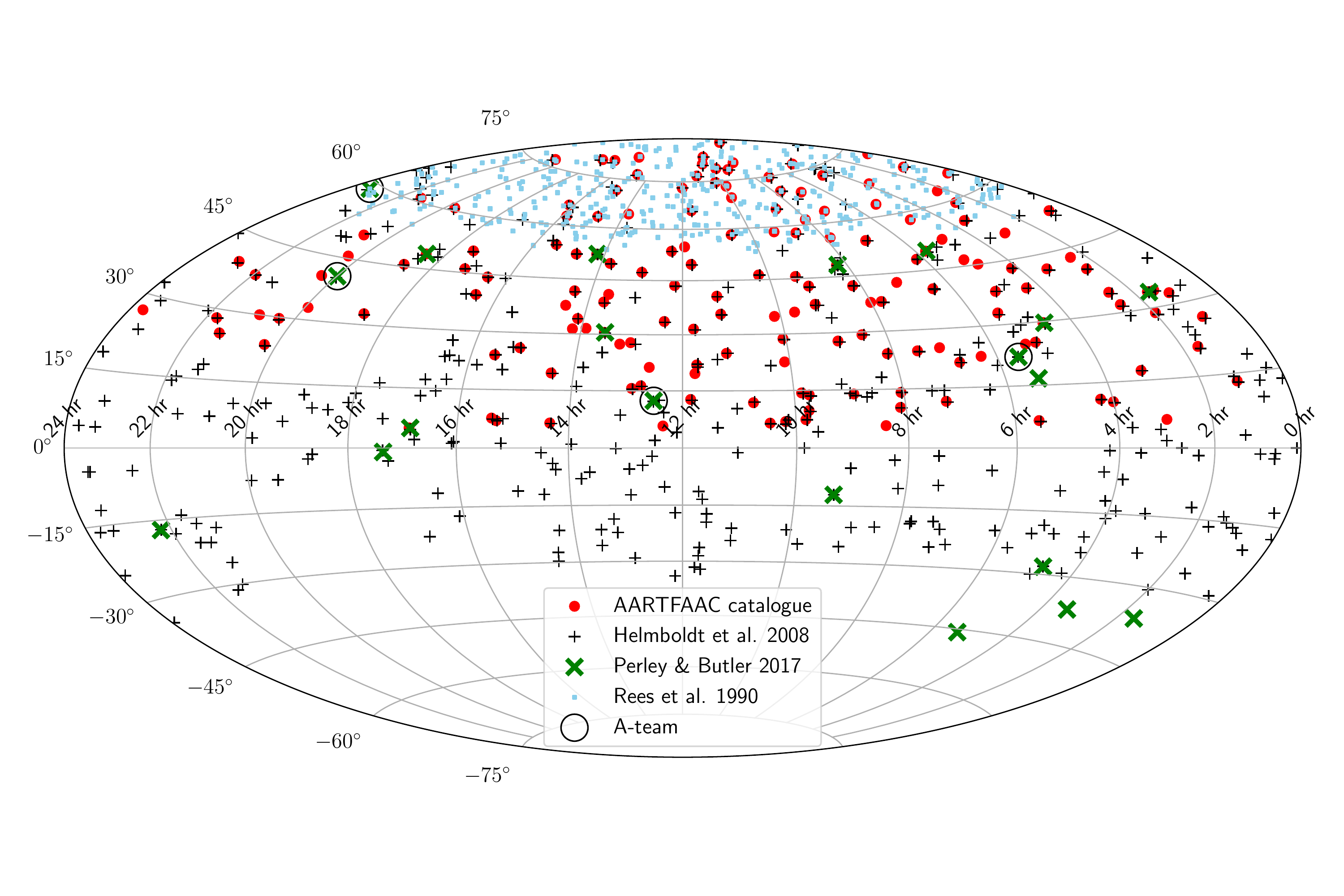}
\caption{
A comparison of the 167 AARTFAAC catalogue sources, and the primary reference sources:  23 in \protect\cite{2017ApJS..230....7P} of which 8 are observable above a declination of $0^{\circ}$, and 388 from  \protect\cite{2008ApJS..174..313H} of which 120 are associated with AARTFAAC sources.
}
\label{fig:catalog_map}
\end{center}
\end{figure*}

The aim of this catalogue is to fill the gap between the VLSSr at 74 MHz \citep{2014MNRAS.440..327L} and the 8C catalogue at 38 MHz. Indeed, for many sources below the $60^{\circ}$ degree declination limit of the 8C survey, this catalogue contains the lowest frequency flux density measurement. The source detection region extends to $50^{\circ}$ from zenith. As a result of correcting the effect of the primary beam, the background noise increases from zenith toward the horizon. However,  within $50^{\circ}$ from zenith the noise is roughly constant or increases slowly. Beyond $50^{\circ}$ however, the background noise is greater than twice the noise at zenith and increases dramatically. Given that zenith is toward $52.9^{\circ}$ declination at the LOFAR superterp, the survey covers the full Northern Hemisphere. This is illustrated in Fig. \ref{fig:catalog_map}, where the coverage area of the AARTFAAC catalogue is  compared to the \cite{2017ApJS..230....7P}, \cite{2008ApJS..174..313H}, and the 8C \citep{1990MNRAS.244..233R} catalogues.

This catalogue will also be beneficial as an independent check for the calibration of low frequency, wide field images generated by the LWA and other LOFAR-LBA surveys such as the forthcoming MSSS. Additionally, the technique presented here can be implemented for real-time streaming calibration of the Southern Hemisphere arrays MWA and SKA-LOW.

\subsection{Position}

The uncertainties in the position measurements by AARTFAAC are large relative to other surveys due to the lower resolution of the images. Fortunately, given the threshold to which we detect sources, the resulting number density in the sky is such that this does not cause a problem when associating measurements of any source across the set of images in an observation. Correspondingly, within an association radius of $1^{\circ}$ any AARTFAAC source can be uniquely matched. Moreover, since the primary goal is the creation of a catalogue for flux density calibration to compare with future transients, highly accurate source positions are not essential.

Nevertheless, the best estimate of the position of each source was measured. This was done by taking the mean, weighted by the uncertainty in the fitted position from each extraction.

\begin{figure}
\begin{center}
\includegraphics[width=0.5\textwidth]{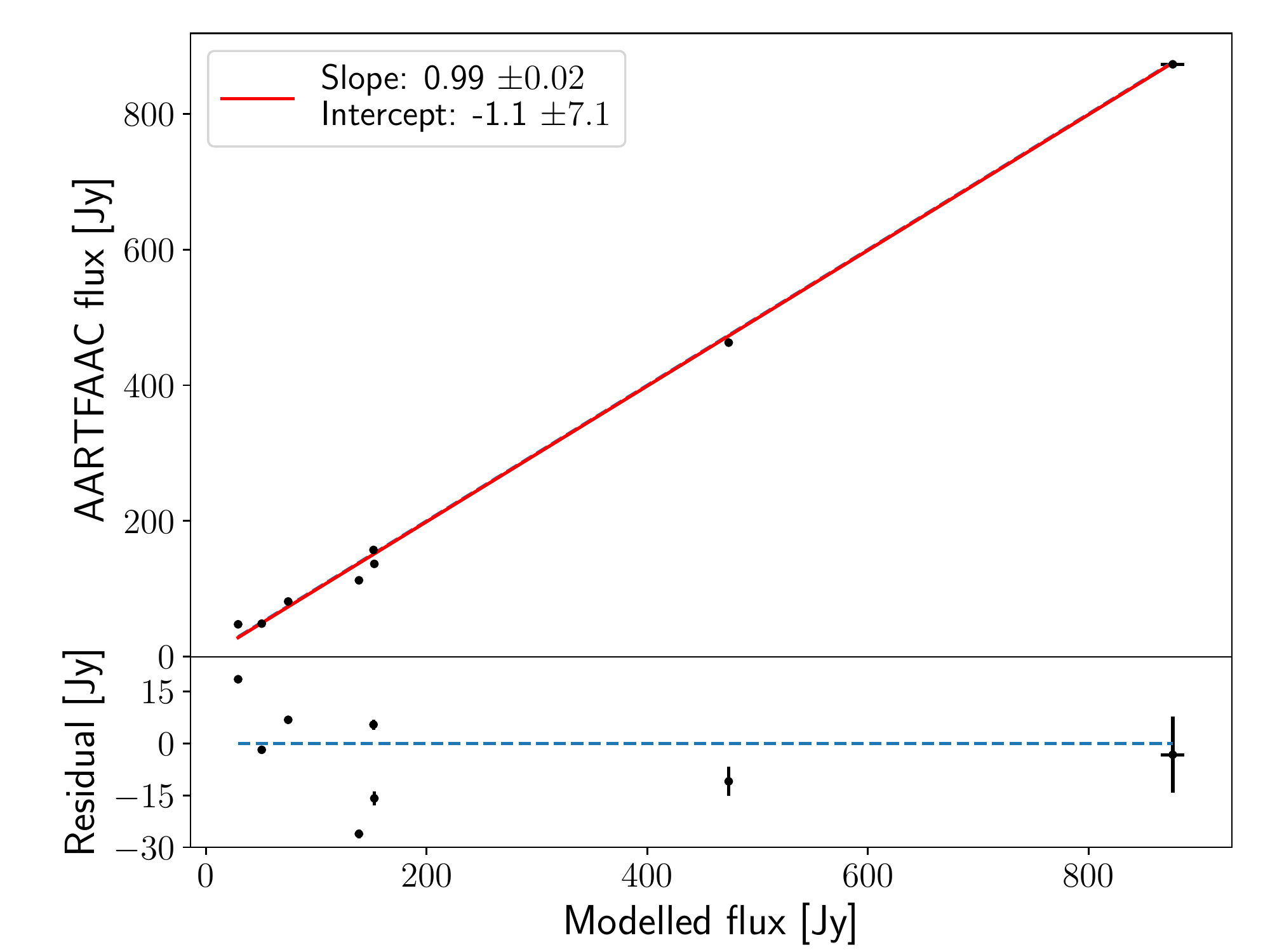}
\caption{Comparing the modelled flux density of the initial calibrators, from \citet{2017ApJS..230....7P} to their AARTFAAC measured flux density. This shows excellent relative brightness agreement across the sources. The exception is the faintest source 3C286, with a modelled flux density of $29.15 \pm 0.25$ Jy and a measured flux density of $47.69 \pm 0.12$ Jy. Measurements reported by \citet{2008ApJS..174..313H} also have a large scatter around the model across the frequency range. This may be due to systematic differences among the different surveys.   } 
\label{fig:catalogcompare} 
\end{center}
\end{figure}

\subsection{Reference source fit}

\begin{table*}
\centering
\begin{tabular}{l|r|r|r|r}
\hline
\hline
 Source Name   &	Scaife and Heald 	& Perley and Butler & AARTFAAC	& Eastwood et al.\\
 			&   Model 					&	 Model   		&  Measured & Interpolated \\ 
			&   [Jy] @ 60 MHz		& [Jy] @ 60 MHz  	& [Jy] @ 60 MHz & [Jy] @ 60 MHz  \\ \hline
3C48 &   	$ 77 \pm 5 $	&	$ 74.55 \pm 0.64 $		&	$ 81.36 \pm 0.33  $ & $ 83.7 \pm 2.29 $	\\
3C123 &   	-				&	$ 473.75 \pm 3.87 $		&	$ 462.85 \pm 0.33 $ &	- \\
3C147 &   	$ 43 \pm 4 $ 	&	$ 50.59 \pm 0.66 $		&	$ 48.71 \pm 0.14 $ & $ 44.89 \pm 1.02 $ \\
3C196 &   	$ 151 \pm 5$	&	$ 151.84 \pm 1.25 $  	&	$ 157.30 \pm 0.20 $	& - \\
3C286 &  	$ 33 \pm 2 $	&	$ 29.15 \pm 0.25 $		&	$ 47.69 \pm 0.12 $ & $ 32.63 \pm 0.43 $	\\
3C295 &  	$ 134 \pm 11$	&	$ 138.65 \pm 1.13 $ 	&	$ 112.55 \pm 0.18 $ & $ 125.22 \pm 3.44 $	\\
3C380 &  	$ 156 \pm 4 $	& 	$ 152.63 \pm 1.95 $ 	&	$ 136.75 \pm 0.09 $	& $ 134.49 \pm 3.70 $ \\ 
Hercules A &	- 	& 	$ 876.31 \pm 10.55  $ 	&	$ 873.07 \pm 0.37 $	& - \\
\hline \hline
\end{tabular}
\caption{A comparison of the difference between the AARTFAAC inferred flux density, and values modelled from the spectral fits presented in the reference catalogues. The modelled flux density, and associated uncertainties, for the \citet{2017ApJS..230....7P} catalogue were generated via a Monte Carlo method, by fitting a spectrum to random samples of the flux density measurements.  These values were used as the initial flux density scale for bootstrapping to the entire AARTFAC catalogue. Additionally for comparison, the modelled flux density from \citet{2012MNRAS.423L..30S} are shown. These flux density estimates, and uncertainties, were generated using the coefficients and their uncertainty in the spectral model published. As such, the uncertainty in these flux density estimates at 60 MHz is much higher. Lastly, the our values are compared to the results of \citet{2018AJ....156...32E}, interpolated to 60 MHz.}
\label{table:flux_compare}
\end{table*}

The accuracy with which the fluxes of the modelled reference sources are measured from the images, after calibration, is a validation of the models. This is due to the fact that, as more sources are added to the catalogue and incorporated into the calibration scheme, the inferred flux density of a single incorrectly modelled source would diverge from the initial estimate, due to the influence of the other correctly modelled sources on the flux density scaling fit for the entire image. 

In order to illustrate the resulting goodness of fit between the modelled reference fluxes used in calibration and the resulting measurements, the spectrum from \citet{2017ApJS..230....7P}, with our AARTFAAC data point (open circle), is shown in Fig. \ref{fig:3c380_spectra}. Here, the AARTFAAC measured flux density does not agree, within the uncertainty of the reference model (blue region). In fact, Fig.  \ref{fig:catalogcompare} and Table \ref{table:flux_compare} show that the only source for which the reference model and AARTFAAC measurement agree within their combined uncertainty is Hercules A, the brightest source.

However, a strong linear relationship between the reference flux density and the measured flux density, is illustrated in Fig. \ref{fig:catalogcompare}. This illustrates the linear response of the array to sources from 50 to over 800 Jy. And suggests that the relative brightnesses of the models are not completely accurate at 60 MHz.

Lastly, our flux density values were compared to those measured by the Owens Valley Long Wavelength Array (OVRO-LWA), \citep{2018AJ....156...32E}. The OVRO-LWA flux density values compared in Table \ref{table:flux_compare} are derived by interpolating between the measured values, provided in their supplementary online material, and the uncertainties calculated using a Monte Carlo method to randomly draw a population of flux density measurements within the reported uncertainty range. Notably, the sources 3C48 and 3C380 both agree within mutual uncertainties. This is interesting due to the similarity of the instruments, but very different method for calibration and imaging.

\begin{figure}
\begin{center}
\includegraphics[width=0.5\textwidth]{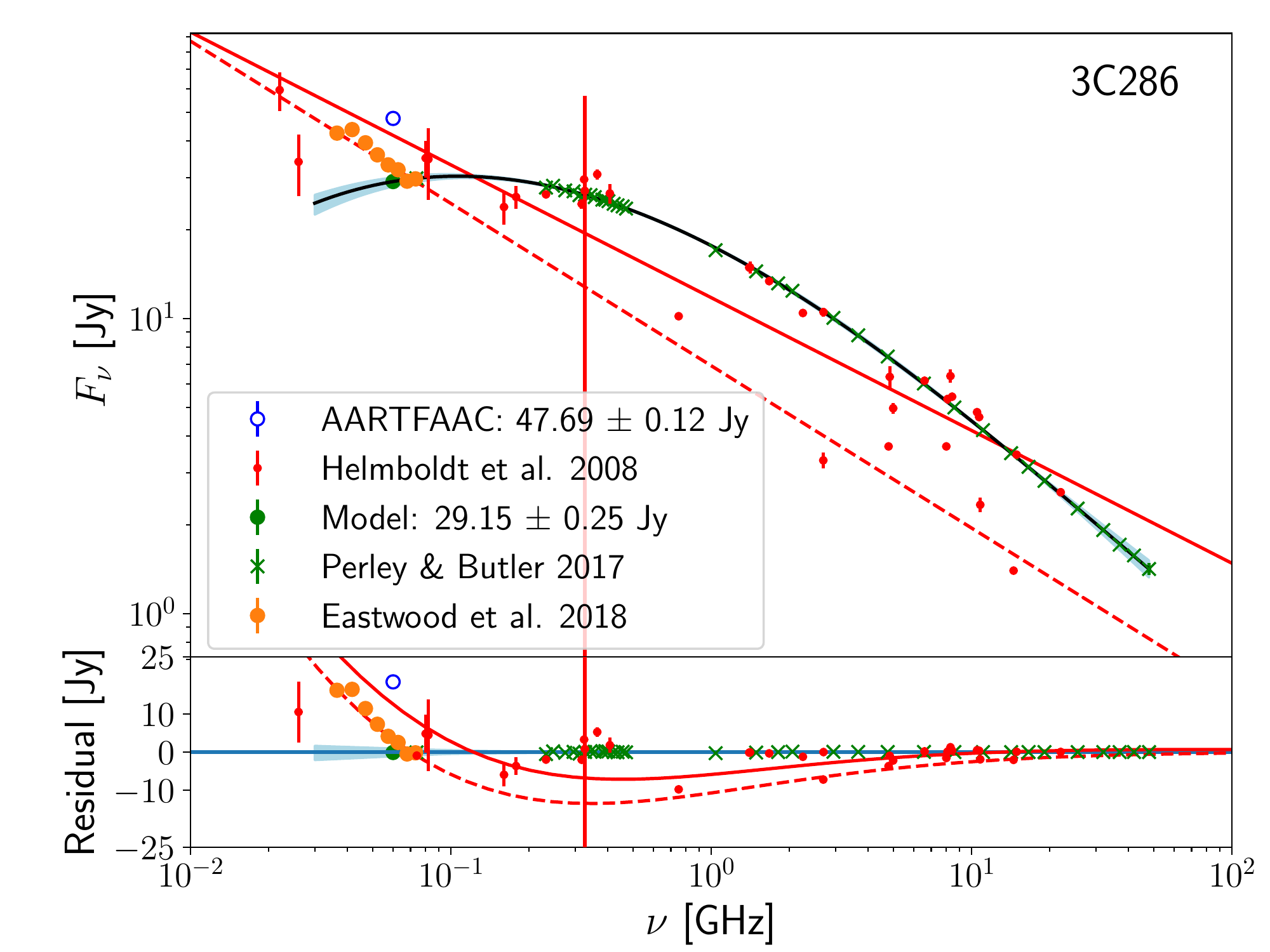}
\caption{A comparison of the collected reference measurements of the flux density of 3C286, compared to the AARTFAAC measurements (open circle).  The \citet{2017ApJS..230....7P} measurements (green x's) and  modelled spectra (black line) was used for the initial catalogue bootstrapping. While values reported by \citet{2018AJ....156...32E} (orange circle) are direct measurements at frequencies comparable to AARTFAAC with the OVRO-LWA, an instrument with a similar design to AARTFAAC. However the flux densities reported by \citet{2008ApJS..174..313H} (red dots) show the high variance in published flux densities across the spectrum. Two simple power-laws (solid and dashed red lines) fit by \citet{2008ApJS..174..313H} are also shown. Interestingly both AARTFAAC and OVRO-LWA agree better with the simple power-laws shown here. 
}
\label{fig:3c286_perley}
\end{center}
\end{figure}

Lastly, it is apparent that the flux density measured here for 3C286 is significantly higher than what is reported by all three of the reference catalogues. It is unclear what could cause this for a single source. 3C286 is a very well known calibrator. In an attempt to explain the additional flux, measurements by a single dish instrument, in which 3C286 is unresolved \citep{1994A&A...284..331O} were compared. However, this study  yielded results that agree with the measurements of \citet{2017ApJS..230....7P} using the VLA, indicating that we are not observing additional diffuse flux as in the case of the Tycho supernova remnant. Additionally, given that 3C286 is at a high Galactic latitude, ~$10^{\circ}$ north of the North Galactic Spur, it is unlike likely the additional flux is the result of diffuse Galactic emission removed by simply imposing the minimum baseline of $10\lambda$. Further, there is no correlation between sky location and deviations between the modelled and measured flux densities. When viewing the AARTFAAC measured flux density alongside the flux density values and simple spectra reported by \citet{2008ApJS..174..313H}, illustrated in Fig. \ref{fig:3c286_perley}, the difference does not appear as anomalous. In fact the power-law spectra fit by \citet{2008ApJS..174..313H} predict a flux density  at 60 MHz of 42 Jy. While these data and spectra are less precise than those measured by \citet{2017ApJS..230....7P}, it is notable that both the results from AARTFAAC and the OVRO-LWA are better predicted by these spectra.

\subsection{Spectral turnovers}

Some of the spectral models fit by \citet{2008ApJS..174..313H}  predict a turnover below the lowest frequency at which measurements were made. Nevertheless, the new flux density measurements made at 60 MHz clearly indicate that, instead, the spectral shape of at least six of these sources continue to rise. The spectra for the six sources, whose labels from both the AARTFAAC catalogue and VLSSr catalogue are listed in Table \ref{tab:turnover}, can be seen in the supplemental online material, where the AARTFAAC measurement is plotted alongside the flux density measurements and spectral fits from \citet{2008ApJS..174..313H}. An example of these figures is shown in Fig. \ref{fig:tycho_spectra}, where the flux density measurements from \citet{1979A&A....76..120K} are plotted as well.

Further ongoing flux density studies, across the full observational spectrum of the LOFAR LBA, 10-90 MHz, could potentially reveal turnovers at a frequency lower than 60 MHz.

\begin{table}
\centering
\begin{tabular}{l|r|r|r}
\hline
\hline
 AARTFAAC   	&  		& VLSSr  \\
	Label		&   	 	&	Label		 	 \\ \hline
J0110+322	&   -	&	J0110+315			\\ 
J1011+068	&   -	&	J1011+062			\\ 
J1147+496	&   -	&	J1146+495			\\ 
J1445+768	&   -	&	J1447+766			\\ 
J1630+442 	&	- 	& 	J1629+442 			\\ \hline \hline
\end{tabular}
\caption{A list of sources for which a  spectral turnover was predicted,  which we do not observe. Here the signifier from the AARTFAAC and VLSSr are given. In the supplemental online material the AARTFAAC flux density measurements are compared alongside the measurements and spectral fits in figure 1 of \citet{2008ApJS..174..313H}. 
}
\label{tab:turnover}
\end{table}

\subsection{Blended sources}

One issue with interferometers is that the minimum baseline length determines the sensitivity of the instrument to regions with large diffuse emission. Therefore, objects with a larger angular size will have their diffuse emission, at least partially, resolved out by interferometers which achieve a high angular resolution. This reduces the total apparent flux density of diffuse sources when compared to measurements by a single dish instrument. Consequently, in order to remove the large, bright, diffuse structure of the Milky Way we eliminate all baselines below $10 \lambda$. This effectively eliminates much of the Galactic emission which would otherwise be a dominant foreground.

\begin{figure}
\begin{center}
\includegraphics[width=0.5\textwidth]{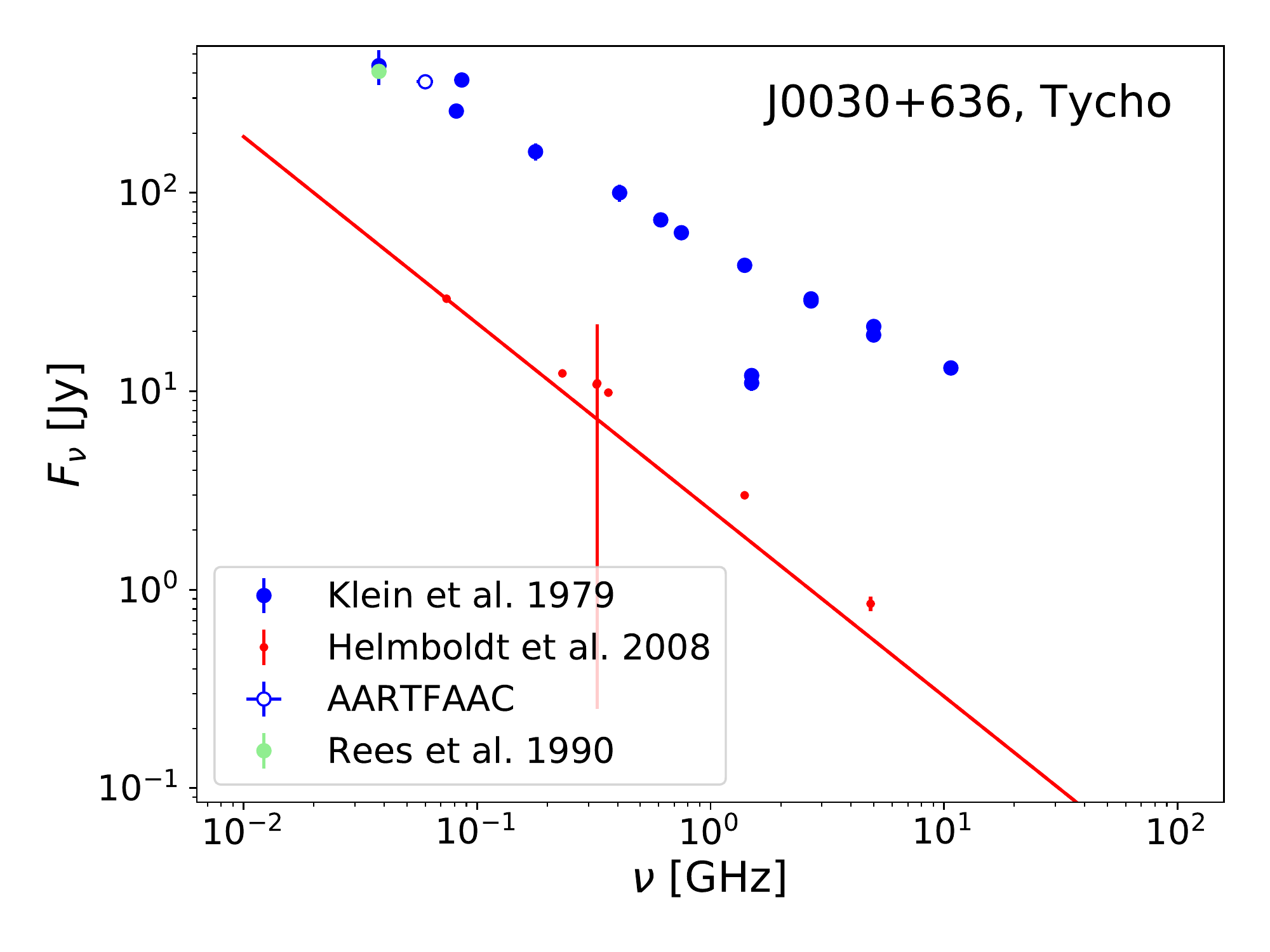}
\caption{The spectra of the supernova remnant 1572, also known as Tycho. The measurements reported by \citet{2008ApJS..174..313H} (red dots) show a significantly reduced integrated flux density compared to the AARTFAAC measurement at 60 MHz. This is due to the higher resolution of these surveys resolving out power in the diffuse emission. Comparing this to the multi-frequency single dish measurements by \citet{1979A&A....76..120K} (blue circles), the 8C catalogue  \citep{1990MNRAS.244..233R} (green circle), and AARTFAAC (open circle) which fully recover flux density on this scale like a single dish instrument, as a result of  their lower resolution. 
}
\label{fig:tycho_spectra}
\end{center}
\end{figure}

In observing large diffuse sources, such as nearby supernova remnants, we see that AARTFAAC recovers the total integrated flux density as effectively as a single dish instrument. This effect is shown in Fig. \ref{fig:tycho_spectra} for the supernova remnant Tycho, where the flux density measurements (red dots) and spectral fit (red line) reported by \mbox{\citet{2008ApJS..174..313H}} are compared with a multi-wavelength analysis observed with a single dish instrument by \mbox{\citet{1979A&A....76..120K}}(blue circles). Additionally, the integrated flux density value from the 8C catalogue is plotted in green. Consequently, the sensitivity of AARTFAAC on larger angular scales presents an additional use case beyond transients; for example, to study Galactic emission, and diffuse regions around other sources such as Cassiopeia A and Cygnus A.

Due to the relatively low resolution of AARTFAAC, compared to the 8C and VLSSr surveys, some objects reported here as individual sources are known to be composed of two or more sources blended together. By comparing the AARTFAAC catalogue to VLSSr with integrated flux density greater than 10 Jy we find that the sources listed in Table \ref{tab:blended}
are the result of multiple blended objects.

\begin{table}
\centering
\begin{tabular}{l|r|r}
\hline
\hline
 AARTFAAC   	& VLSSr fluxes & AARTFAAC flux density 					\\
	Label		&   [Jy]	 	&	[Jy]						 \\ \hline
J0027+643 	&   $ 28.24 \pm 0.19 $	&	$ 362.42 \pm 0.27 $				\\
- 		  	&   $ 17.62 \pm 0.19 $	&	-								\\
- 		  	&   $ 13.51 \pm 0.20 $	&	-								\\
J0110+134 	&   $ 56.78 \pm 0.10 $	&	$ 122.93 \pm 0.23 $				\\
- 		  	&   $ 23.69 \pm 0.10 $	&	-								\\
J0224+430 	&   $ 12.77 \pm 0.15 $	&	$ 96.47 \pm 0.10 $				\\
- 		  	&   $ 10.16 \pm 0.16 $	&	-								\\
J0400+105 	&   $ 23.88 \pm 0.26 $	&	$ 132.39 \pm 0.15 $				\\
- 		  	&   $ 10.85 \pm 0.26 $	&	-								\\
J0420+381 	&   $ 38.76 \pm 0.21 $	&	$ 167.93 \pm 0.09 $				\\
- 		  	&   $ 29.24 \pm 0.21 $	&	-								\\
- 		  	&   $ 13.44 \pm 0.21 $	&	-								\\
J0506+381 	&   $ 86.76 \pm 0.24 $	&	$ 237.15 \pm 0.13 $				\\
- 		  	&   $ 25.40 \pm 0.24 $	&	-								\\
- 		  	&   $ 14.56 \pm 0.24 $	&	-								\\
- 		  	&   $ 11.31 \pm 0.24 $	&	-								\\
J0627+401 	&   $ 16.76 \pm 0.10 $	&	$ 44.95 \pm 0.06 $				\\
- 		  	&   $ 16.31 \pm 0.09 $	&	-								\\
J0657+542 	&   $ 39.71 \pm 0.09 $	&	$  56.75 \pm 0.06 $				\\
- 		  	&   $ 12.61 \pm 0.10 $	&	-								\\
J0829+292 	&   $ 19.55 \pm 0.10 $	&	$ 39.14 \pm 0.08 $				\\
- 		  	&   $ 15.38 \pm 0.10 $	&	-								\\
J0855+139 	&   $ 40.21 \pm 0.17 $	&	$ 64.32 \pm 0.13 $				\\
- 		  	&   $ 15.17 \pm 0.18 $	&	-								\\
J1144+218 	&   $ 30.70 \pm 0.14 $	&	$ 50.02 \pm 0.16 $				\\
- 		  	&   $ 16.40 \pm 0.13 $	&	-								\\
J1445+768 	&   $ 18.67 \pm 0.13 $	&	$ 45.58 \pm 0.10 $				\\
- 		  	&   $ 14.72 \pm 0.14 $	&	-								\\
J1506+259 	&   $ 28.97 \pm 0.12 $	&	$ 127.36 \pm 0.13 $				\\
- 		  	&   $ 11.99 \pm 0.12 $	&	-								\\
J1552+050 	&   $ 373.91 \pm 0.49 $	&	$ 873.07 \pm 0.37 $				\\
- 		  	&   $ 309.02 \pm 0.51 $	&	-								\\
J1840+797 	&   $ 22.01 \pm 0.32 $	&	$ 117.25 \pm 0.14 $				\\
- 		  	&   $ 18.60 \pm 0.31 $	&	-								\\
- 		  	&   $ 10.92 \pm 0.31 $	&	-								\\
J2247+397 	&   $ 34.83 \pm 0.18 $	&	$ 150.89 \pm 0.15 $				\\
- 		  	&   $ 25.58 \pm 0.19 $	&	-								\\
- 		  	&   $ 25.46 \pm 0.18 $	&	-								\\
\hline \hline
\end{tabular}
\caption{A list of sources in the AARTFAAC catalogue which are known to be composed of two or more VLSSr sources, blended together. A threshold of 10 Jy was used to filter the VLSSr catalogue in order to limit the number of objects compared. This threshold is motivated by the fact that a source below 10 Jy at 74 MHZ would, given the sensitivity of AARTFAAC, likely not have a strong contribution to the observed flux density.  }
\label{tab:blended}
\end{table}

However, so long as the sum of these blended sources maintains a stable flux density the component contribution of each is not important for us, since AARTFAAC does not resolve them independently. Therefore, no effort was made to de-blend the sources, and attempt to retrieve an accurate flux density for each individually. The blended sources, as they appear to AARTFAAC, are still useful for our calibration purposes. Although instruments with a higher resolution, including AARTFAAC after the currently planned extension which will incorporate 6 additional stations, may need to exclude these when calibrating, and measure the separate component fluxes.  
 
\subsection{Summary of flux calibration and measurement accuracy}

From all the above, it is clear that a number of significant factors
play a role in the accuracy with which AARTFAAC can report calibrated
source fluxes: the stability and sensitivity of our instrument, the
stability of the ionosphere, and the ability to relate our
instrumental brightness measurements to well studied flux calibrators.
Here we collect and summarize our findings on these aspects:

Instrumental flux stability: We collected data from all LSTs,
over a 3-month period, and find that longer-time average fluxes show
no trends with time or Zenith angle at levels above 1\%, so indeed our
measurements and instrumental calibration are quite stable
(Figs.~\ref{fig:beamexample} and \ref{fig:tycho_noise}). Furthermore, we see
that for bright sources the error estimates on 1-second measurements are
a bit larger than the measured flux variations on short time scales
(Fig.~\ref{fig:tycho_noise}), so our error estimates are
somewhat conservative.

Ionospheric effects: We have used data from representative
ionospheric weather, however excluding periods of either strong RFI or ionospheric turbulence around the A-team sources which resulted the data being flagged by the correlator  or calibration pipelines, and no images being created. We
see that in these data, the dominant timescale for ionospheric
variations is of order 15 minutes at our observing frequency (60\,MHz)
and the typical amplitude is 10--15\%, somewhat larger than the
instantaneous flux measurement accuracy of bright sources
(Fig.~\ref{fig:tycho_noise}).  We also find that these variations are
spatially uncorrelated on scales more than 5 degrees. This is why we
employ the strategy of fitting instantaneous flux scales using all
available AARTFAAC catalogue sources at any time: it decreases the
uncertainty in the calibration scale factor and prevents the scintillation variations of a single calibrator to
dominate the flux scale. Because we do this every second and monitor
the variations, we are provided with an automatic assessment of
ionospheric conditions, which is also used by LOFAR. These effects
are much less in MWA at somewhat higher frequencies, see
\citep{2015MNRAS.453.2731L}, indicating that even within the LOFAR low band
the
strength will vary significantly with frequency.

Absolute calibration: To tie the AARTFAAC flux scale to 
more widely applicable radio flux
calibrations, we compared our fluxes to a number of previously
established radio catalogues. We had to fit models and interpolate,
since very few previous measurements are available at 60\,MHz
(sect.~\ref{sec:cat}).  In table~\ref{table:flux_compare}, we can see the precision and
stability of our measurements is indeed very good compared to previous
work, but that the calibrations of different very bright Northern
sources differ by a few to 10\% between papers, and in a few cases
more (specifically our flux for 3C286 seems anomalously high compared
to other work).  Absolute flux calibration at these radio frequencies
thus appears to be mostly reliable to 10\%.

\section{Conclusion}
\label{conclusion}

This work presents the method used for calibrating the flux density scale of AARTFAAC images in real-time, for the upcoming transient and variability surveys. Additionally, the AARTFAAC catalogue of calibration sources is presented.  

Due to the lack of a sufficient number of well measured calibrator sources at low frequencies, and the requirement that AARTFAAC images are calibrated in real-time, a bootstrap algorithm was used. Hence, the AARTFAAC catalogue is fundamentally based on a flux density scaling derived from the spectral fits published by \citet{2017ApJS..230....7P}. Consequently, any systematic bias in the fluxes reported there could influence the AARTFAAC catalogue. Therefore, as a verification the AARTFAAC catalogue was compared to the larger, but less precise, catalogue of \citet{2008ApJS..174..313H}. Here good agreement was found between the spectral fit extrapolated to 60 MHz and the measurements presented in this work. However, a tendency for AARTFAAC to measure more flux was observed. This is potentially explained by the much higher density of the LOFAR superterp resulting in higher sensitivity to diffuse emission. Therefore, it is shown that AARTFAAC is capable of filling the gap between 38 and 74 MHz and providing an accurate flux density scale for the calibration of densely packed low-frequency arrays.

Additionally, statistical analysis of the times series of flux density measurements for each source resulted in insights into the effect of ionospheric variability. Significantly, it was observed that such variability preferentially skews the distribution of measurements in the positive direction, giving the average a positive bias. Consequently, it was found that the most robust method to mitigated this was to fit a skew normal distribution, and calculate the mode. In light of this, observations which attempt to achieve high sensitivity with long integrations, without correcting for ionospheric variability on short timescales, will also have a positive bias.

For our ability to detect new transients, the flux calibration is of
minor importance, because this depends only on the ability to detect a
new source against the local noise. The main limitations for this are
(i) that we examine large volumes of data, so we typically have
to set the threshold at 8 times the local RMS to avoid many false
positives, and (ii) that we have to distinguish transients from RFI,
terrestrial sources and artifacts. But if we do detect a source, the
instantaneous flux uncertainty on second time scales will be dominated
by the ionosphere for bright sources, and roughly equally by the
ionospheric and measurement noise near the threshold. The ionospheric
uncertainty decreases significantly only when the sources stay on for
longer than the typical variation time of 15 minutes.

In the future this work may be extended by incorporating more data. Given that the primary science goal of AARTFAAC is to survey the sky for transient and variable sources, many hundreds of hours of observations will be made. This enormous amount of data will allow for extremely precise flux density values to be measured.

Lastly, observations across the entire frequency range of the LOFAR LBA 10-90 MHz would allow highly accurate spectra to be fit, however at the low end of the bandpass \citet{2017ApJS..230....7P} would no longer be a suitable reference catalogue. Lastly, the planned upgrade of AARTFAAC to include 12 stations of the LOFAR core, rather than the six currently in operation, will produce images of higher sensitivity and resolution. This will allow for the measurement of currently blended sources, and many which are now below our detection threshold.

\section*{Acknowledgements}

This work is funded by the ERC Advanced Investigator grant no. 247295 awarded to Prof. Ralph Wijers, University of Amsterdam. We thank The Netherlands Institute  for Radio Astronomy (ASTRON) for support provided in carrying out the commissioning observations.

The authors would also like to thank the LOFAR science support for their assistance in obtaining and processing the data used in this work. We use data obtained from LOFAR, the Low Frequency Array designed and constructed by ASTRON, which has facilities in several countries, that are owned by various parties (each with their own funding sources), and that are collectively operated by the International LOFAR Telescope (ILT) foundation under a joint scientific policy.




\bibliographystyle{mnras.bst}
\bibliography{FluxandCat.bib}  



\onecolumn
\appendix

\begin{landscape}
\section{Supplementary online material: AARTFAAC catalogue}
\label{apx:cat}

\begin{longtable}{lrrrrrrrrl}
\hline \hline 
Label	& Ra     	& Dec    & Flux density    & Location & Scale & Shape & Name & Measurements    \\
      	& [$^{\circ}$]  &  [$^{\circ}$] & [Jy]   & $\xi$ & $\omega$  & $\alpha$  & &  \# \\\hline 
\endfirsthead
\multicolumn{8}{@{}l}{\ldots continued}\\\hline
Label	& Ra     	& Dec    & Flux density    & Location & Scale & Shape & Name & Measurements    \\
      	& [$^{\circ}$]  &  [$^{\circ}$] & [Jy]   & $\xi$ & $\omega$  & $\alpha$ &  & \# \\
\hline
\endhead
\hline 
\multicolumn{8}{r@{}}{continued \ldots} \\    
\endfoot
\endlastfoot
J0007+725 & $ 1.96 \pm 0.36 $ & $ 72.53 \pm 0.12 $ & $ 86.17 \pm 0.13 $ & $ 75.94 \pm 0.09 $ & $ 22.17 \pm 0.09 $ & $ 3.27 \pm 0.05 $ &  CTA 1 & $ 72070$ \\ 
J0027+642 & $ 6.78 \pm 0.10 $ & $ 64.22 \pm 0.10 $ & $ 362.42 \pm 0.27 $ & $ 336.64 \pm 0.31 $ & $ 48.75 \pm 0.27 $ & $ 2.10 \pm 0.04 $ &  Tycho SNR & $ 55465$ \\ 
J0044+521 & $ 11.19 \pm 0.05 $ & $ 52.20 \pm 0.11 $ & $ 70.75 \pm 0.09 $ & $ 63.80 \pm 0.08 $ & $ 14.30 \pm 0.08 $ & $ 2.86 \pm 0.05 $ &  3C20 & $ 35940$ \\ 
J0057+266 & $ 14.45 \pm 0.06 $ & $ 26.66 \pm 0.09 $ & $ 72.75 \pm 0.14 $ & $ 65.78 \pm 0.30 $ & $ 13.20 \pm 0.20 $ & $ 1.17 \pm 0.06 $ &  3C28 & $ 18137$ \\ 
J0058+683 & $ 14.64 \pm 0.20 $ & $ 68.39 \pm 0.11 $ & $ 91.20 \pm 0.08 $ & $ 85.74 \pm 0.26 $ & $ 11.45 \pm 0.14 $ & $ 0.86 \pm 0.05 $ &  3C27 & $ 41301$ \\ 
J0110+322 & $ 17.54 \pm 0.09 $ & $ 32.26 \pm 0.11 $ & $ 61.68 \pm 0.12 $ & $ 55.01 \pm 0.12 $ & $ 12.99 \pm 0.11 $ & $ 2.39 \pm 0.07 $ &  3C34 & $ 19807$ \\ 
J0110+134 & $ 17.59 \pm 0.03 $ & $ 13.45 \pm 0.09 $ & $ 122.93 \pm 0.23 $ & $ 111.74 \pm 0.34 $ & $ 20.57 \pm 0.27 $ & $ 1.58 \pm 0.06 $ &  3C33 & $ 11207$ \\ 
J0127+332 & $ 21.97 \pm 0.12 $ & $ 33.30 \pm 0.13 $ & $ 51.72 \pm 0.14 $ & $ 47.55 \pm 0.22 $ & $ 7.68 \pm 0.17 $ & $ 1.54 \pm 0.11 $ &  3C41 & $ 5048$ \\ 
J0137+210 & $ 24.47 \pm 0.04 $ & $ 21.08 \pm 0.07 $ & $ 89.87 \pm 0.18 $ & $ 79.84 \pm 0.23 $ & $ 18.60 \pm 0.19 $ & $ 1.83 \pm 0.06 $ &  3C47 & $ 16127$ \\ 
J0139+332 & $ 24.84 \pm 0.04 $ & $ 33.29 \pm 0.10 $ & $ 81.36 \pm 0.33 $ & $ 62.62 \pm 0.26 $ & $ 39.52 \pm 0.26 $ & $ 3.05 \pm 0.07 $ &  3C48 & $ 25907$ \\ 
J0158+289 & $ 29.65 \pm 0.05 $ & $ 28.96 \pm 0.09 $ & $ 48.71 \pm 0.14 $ & $ 42.46 \pm 0.14 $ & $ 12.04 \pm 0.13 $ & $ 2.28 \pm 0.08 $ &  3C55 & $ 14031$ \\ 
J0220+625 & $ 35.09 \pm 0.09 $ & $ 62.54 \pm 0.12 $ & $ 158.57 \pm 0.18 $ & $ 142.35 \pm 0.16 $ & $ 33.87 \pm 0.16 $ & $ 2.97 \pm 0.05 $ &  HB3 & $ 48715$ \\ 
J0224+430 & $ 36.13 \pm 0.05 $ & $ 43.06 \pm 0.10 $ & $ 96.47 \pm 0.10 $ & $ 88.40 \pm 0.09 $ & $ 17.03 \pm 0.09 $ & $ 3.06 \pm 0.05 $ &  3C66 & $ 36406$ \\ 
J0224+401 & $ 36.19 \pm 0.07 $ & $ 40.15 \pm 0.07 $ & $ 55.38 \pm 0.08 $ & $ 49.47 \pm 0.06 $ & $ 13.53 \pm 0.08 $ & $ 3.72 \pm 0.07 $ &  3C65 & $ 27047$ \\ 
J0225+863 & $ 36.30 \pm 1.41 $ & $ 86.36 \pm 0.09 $ & $ 100.86 \pm 0.09 $ & $ 108.41 \pm 0.37 $ & $ 16.51 \pm 0.18 $ & $ -0.78 \pm 0.04 $ &  - & $ 121716$ \\ 
J0234+346 & $ 38.51 \pm 0.06 $ & $ 34.60 \pm 0.09 $ & $ 45.81 \pm 0.10 $ & $ 41.18 \pm 0.09 $ & $ 9.28 \pm 0.09 $ & $ 2.63 \pm 0.08 $ &  - & $ 14837$ \\ 
J0235+317 & $ 38.96 \pm 0.06 $ & $ 31.71 \pm 0.06 $ & $ 45.11 \pm 0.14 $ & $ 40.65 \pm 0.22 $ & $ 8.22 \pm 0.16 $ & $ 1.48 \pm 0.10 $ &  - & $ 7556$ \\ 
J0240+593 & $ 40.10 \pm 0.08 $ & $ 59.37 \pm 0.06 $ & $ 56.82 \pm 0.08 $ & $ 52.41 \pm 0.09 $ & $ 8.43 \pm 0.08 $ & $ 2.19 \pm 0.06 $ &  3C69 & $ 18542$ \\ 
J0258+062 & $ 44.71 \pm 0.03 $ & $ 6.23 \pm 0.06 $ & $ 80.23 \pm 0.21 $ & $ 74.50 \pm 0.59 $ & $ 11.38 \pm 0.34 $ & $ 1.00 \pm 0.12 $ &  3C75 & $ 7650$ \\ 
J0302+508 & $ 45.65 \pm 0.09 $ & $ 50.87 \pm 0.07 $ & $ 50.00 \pm 0.09 $ & $ 45.79 \pm 0.15 $ & $ 7.78 \pm 0.11 $ & $ 1.41 \pm 0.07 $ &  3C76 & $ 10737$ \\ 
J0311+171 & $ 47.85 \pm 0.03 $ & $ 17.19 \pm 0.06 $ & $ 68.23 \pm 0.10 $ & $ 62.32 \pm 0.16 $ & $ 10.87 \pm 0.12 $ & $ 1.49 \pm 0.05 $ &  3C79 & $ 20126$ \\ 
J0321+416 & $ 50.29 \pm 0.02 $ & $ 41.69 \pm 0.08 $ & $ 306.40 \pm 0.12 $ & $ 295.88 \pm 0.19 $ & $ 19.35 \pm 0.15 $ & $ 1.49 \pm 0.04 $ &  3C84 & $ 34840$ \\ 
J0329+553 & $ 52.43 \pm 0.06 $ & $ 55.38 \pm 0.08 $ & $ 84.62 \pm 0.07 $ & $ 78.45 \pm 0.11 $ & $ 11.36 \pm 0.09 $ & $ 1.44 \pm 0.03 $ &  3C86 & $ 41894$ \\ 
J0349+680 & $ 57.27 \pm 0.17 $ & $ 68.10 \pm 0.08 $ & $ 48.34 \pm 0.06 $ & $ 47.26 \pm 1.02 $ & $ 5.11 \pm 0.19 $ & $ 0.30 \pm 0.28 $ &  - & $ 16358$ \\ 
J0400+105 & $ 60.09 \pm 0.02 $ & $ 10.53 \pm 0.05 $ & $ 132.39 \pm 0.15 $ & $ 125.67 \pm 0.31 $ & $ 12.73 \pm 0.20 $ & $ 1.18 \pm 0.07 $ &  3C98 & $ 18238$ \\ 
J0409+430 & $ 62.46 \pm 0.04 $ & $ 43.09 \pm 0.08 $ & $ 76.12 \pm 0.09 $ & $ 69.52 \pm 0.13 $ & $ 12.13 \pm 0.10 $ & $ 1.53 \pm 0.04 $ &  3C103 & $ 36700$ \\ 
J0415+112 & $ 63.77 \pm 0.03 $ & $ 11.21 \pm 0.07 $ & $ 69.68 \pm 0.06 $ & $ 70.33 \pm 1.43 $ & $ 6.60 \pm 0.19 $ & $ -0.14 \pm 0.29 $ &  3C109 & $ 16943$ \\ 
J0420+381 & $ 65.04 \pm 0.02 $ & $ 38.14 \pm 0.08 $ & $ 167.93 \pm 0.09 $ & $ 161.74 \pm 0.30 $ & $ 12.90 \pm 0.16 $ & $ 0.87 \pm 0.05 $ &  3C111 & $ 35165$ \\ 
J0428+769 & $ 67.22 \pm 0.72 $ & $ 76.94 \pm 0.16 $ & $ 46.53 \pm 0.12 $ & $ 41.60 \pm 0.07 $ & $ 14.78 \pm 0.09 $ & $ 6.08 \pm 0.16 $ &  - & $ 19722$ \\ 
J0438+297 & $ 69.67 \pm 0.02 $ & $ 29.75 \pm 0.07 $ & $ 462.85 \pm 0.33 $ & $ 438.11 \pm 0.49 $ & $ 45.47 \pm 0.37 $ & $ 1.59 \pm 0.04 $ &  3C123 & $ 33194$ \\ 
J0438+726 & $ 69.68 \pm 0.22 $ & $ 72.62 \pm 0.08 $ & $ 45.49 \pm 0.09 $ & $ 41.83 \pm 0.13 $ & $ 6.74 \pm 0.10 $ & $ 1.64 \pm 0.08 $ &  - & $ 10510$ \\ 
J0450+450 & $ 72.66 \pm 0.03 $ & $ 45.08 \pm 0.07 $ & $ 106.13 \pm 0.08 $ & $ 100.43 \pm 0.19 $ & $ 11.00 \pm 0.12 $ & $ 1.08 \pm 0.04 $ &  3C129 & $ 39028$ \\ 
J0455+521 & $ 73.80 \pm 0.06 $ & $ 52.14 \pm 0.07 $ & $ 57.49 \pm 0.06 $ & $ 52.70 \pm 0.07 $ & $ 8.96 \pm 0.06 $ & $ 1.94 \pm 0.04 $ &  3C130 & $ 35922$ \\ 
J0456+579 & $ 74.09 \pm 0.10 $ & $ 58.00 \pm 0.06 $ & $ 43.06 \pm 0.09 $ & $ 40.35 \pm 0.12 $ & $ 5.01 \pm 0.09 $ & $ 1.75 \pm 0.10 $ &  - & $ 6789$ \\ 
J0503+465 & $ 75.81 \pm 0.06 $ & $ 46.56 \pm 0.09 $ & $ 143.52 \pm 0.10 $ & $ 133.65 \pm 0.11 $ & $ 19.15 \pm 0.10 $ & $ 2.35 \pm 0.04 $ &  HB9 & $ 40246$ \\ 
J0504+252 & $ 76.12 \pm 0.04 $ & $ 25.28 \pm 0.08 $ & $ 56.06 \pm 0.17 $ & $ 50.56 \pm 0.18 $ & $ 10.56 \pm 0.17 $ & $ 2.25 \pm 0.11 $ &  3C133 & $ 6752$ \\ 
J0506+381 & $ 76.61 \pm 0.02 $ & $ 38.16 \pm 0.07 $ & $ 237.15 \pm 0.13 $ & $ 226.26 \pm 0.20 $ & $ 20.02 \pm 0.16 $ & $ 1.50 \pm 0.04 $ &  3C134 & $ 37471$ \\ 
J0507+631 & $ 76.90 \pm 0.09 $ & $ 63.13 \pm 0.08 $ & $ 39.66 \pm 0.09 $ & $ 36.57 \pm 0.09 $ & $ 6.08 \pm 0.08 $ & $ 2.47 \pm 0.10 $ &  - & $ 7826$ \\ 
J0518+249 & $ 79.52 \pm 0.04 $ & $ 24.98 \pm 0.06 $ & $ 59.88 \pm 0.09 $ & $ 55.79 \pm 0.15 $ & $ 7.57 \pm 0.10 $ & $ 1.38 \pm 0.06 $ &  - & $ 12926$ \\ 
J0527+329 & $ 81.79 \pm 0.19 $ & $ 32.94 \pm 0.10 $ & $ 56.77 \pm 0.12 $ & $ 52.14 \pm 0.15 $ & $ 8.61 \pm 0.12 $ & $ 1.90 \pm 0.09 $ &  3C141 & $ 10846$ \\ 
J0532+065 & $ 83.23 \pm 0.04 $ & $ 6.56 \pm 0.06 $ & $ 60.90 \pm 0.16 $ & $ 54.45 \pm 0.24 $ & $ 11.86 \pm 0.17 $ & $ 1.58 \pm 0.08 $ &  - & $ 15398$ \\ 
J0544+498 & $ 86.08 \pm 0.06 $ & $ 49.82 \pm 0.07 $ & $ 48.71 \pm 0.14 $ & $ 41.63 \pm 0.10 $ & $ 15.96 \pm 0.10 $ & $ 3.59 \pm 0.10 $ &  3C147 & $ 26781$ \\ 
J0611+480 & $ 92.85 \pm 0.08 $ & $ 48.05 \pm 0.07 $ & $ 46.02 \pm 0.05 $ & $ 42.15 \pm 0.10 $ & $ 7.17 \pm 0.07 $ & $ 1.33 \pm 0.04 $ &  3C153 & $ 31415$ \\ 
J0618+225 & $ 94.70 \pm 0.02 $ & $ 22.59 \pm 0.07 $ & $ 467.75 \pm 0.16 $ & $ 454.06 \pm 0.25 $ & $ 25.17 \pm 0.20 $ & $ 1.52 \pm 0.04 $ &  IC443, 3C157 & $ 33454$ \\ 
J0627+401 & $ 96.76 \pm 0.05 $ & $ 40.20 \pm 0.06 $ & $ 44.95 \pm 0.06 $ & $ 40.34 \pm 0.08 $ & $ 8.63 \pm 0.06 $ & $ 1.99 \pm 0.05 $ &  3C159 & $ 32863$ \\ 
J0646+213 & $ 101.69 \pm 0.05 $ & $ 21.35 \pm 0.08 $ & $ 61.08 \pm 0.12 $ & $ 55.26 \pm 0.17 $ & $ 10.70 \pm 0.13 $ & $ 1.61 \pm 0.06 $ &  3C166 & $ 15903$ \\ 
J0657+693 & $ 104.27 \pm 0.22 $ & $ 69.31 \pm 0.07 $ & $ 37.17 \pm 0.08 $ & $ 34.50 \pm 0.12 $ & $ 4.91 \pm 0.09 $ & $ 1.65 \pm 0.10 $ &  3C169 & $ 8040$ \\ 
J0657+542 & $ 104.47 \pm 0.09 $ & $ 54.23 \pm 0.09 $ & $ 56.75 \pm 0.06 $ & $ 51.24 \pm 0.11 $ & $ 10.21 \pm 0.08 $ & $ 1.34 \pm 0.03 $ &  3C171 & $ 55924$ \\ 
J0703+251 & $ 105.89 \pm 0.05 $ & $ 25.19 \pm 0.07 $ & $ 48.29 \pm 0.08 $ & $ 43.40 \pm 0.10 $ & $ 9.04 \pm 0.08 $ & $ 1.79 \pm 0.05 $ &  3C172 & $ 26701$ \\ 
J0707+633 & $ 106.85 \pm 0.12 $ & $ 63.32 \pm 0.07 $ & $ 39.24 \pm 0.05 $ & $ 35.81 \pm 0.09 $ & $ 6.31 \pm 0.07 $ & $ 1.46 \pm 0.05 $ &  - & $ 26647$ \\ 
J0709+425 & $ 107.26 \pm 0.08 $ & $ 42.53 \pm 0.09 $ & $ 43.76 \pm 0.06 $ & $ 40.37 \pm 0.11 $ & $ 6.33 \pm 0.07 $ & $ 1.27 \pm 0.05 $ &  - & $ 27363$ \\ 
J0712+748 & $ 108.06 \pm 0.29 $ & $ 74.83 \pm 0.10 $ & $ 53.18 \pm 0.06 $ & $ 47.56 \pm 0.10 $ & $ 10.33 \pm 0.07 $ & $ 1.47 \pm 0.03 $ &  - & $ 74465$ \\ 
J0714+116 & $ 108.62 \pm 0.04 $ & $ 11.67 \pm 0.08 $ & $ 73.94 \pm 0.13 $ & $ 67.27 \pm 0.28 $ & $ 12.52 \pm 0.18 $ & $ 1.23 \pm 0.06 $ &  3C175 & $ 22802$ \\ 
J0731+246 & $ 112.75 \pm 0.05 $ & $ 24.61 \pm 0.07 $ & $ 41.16 \pm 0.08 $ & $ 37.80 \pm 0.12 $ & $ 6.18 \pm 0.09 $ & $ 1.50 \pm 0.07 $ &  - & $ 15365$ \\ 
J0742+702 & $ 115.69 \pm 0.15 $ & $ 70.30 \pm 0.07 $ & $ 39.44 \pm 0.08 $ & $ 34.67 \pm 0.07 $ & $ 9.52 \pm 0.07 $ & $ 2.60 \pm 0.06 $ &  3C184 & $ 35689$ \\ 
J0746+377 & $ 116.54 \pm 0.05 $ & $ 37.79 \pm 0.08 $ & $ 49.35 \pm 0.09 $ & $ 42.72 \pm 0.08 $ & $ 13.42 \pm 0.08 $ & $ 2.72 \pm 0.05 $ &  3C186 & $ 39665$ \\ 
J0747+804 & $ 116.91 \pm 0.35 $ & $ 80.40 \pm 0.08 $ & $ 46.89 \pm 0.06 $ & $ 42.43 \pm 0.08 $ & $ 8.27 \pm 0.06 $ & $ 1.83 \pm 0.05 $ &  - & $ 43127$ \\ 
J0750+558 & $ 117.74 \pm 0.07 $ & $ 55.84 \pm 0.07 $ & $ 54.34 \pm 0.05 $ & $ 50.14 \pm 0.13 $ & $ 8.13 \pm 0.08 $ & $ 1.06 \pm 0.04 $ &  - & $ 57674$ \\ 
J0800+377 & $ 120.10 \pm 0.06 $ & $ 37.70 \pm 0.10 $ & $ 41.59 \pm 0.08 $ & $ 37.57 \pm 0.08 $ & $ 7.80 \pm 0.08 $ & $ 2.36 \pm 0.07 $ &  3C189 & $ 14698$ \\ 
J0802+141 & $ 120.73 \pm 0.04 $ & $ 14.13 \pm 0.08 $ & $ 50.68 \pm 0.12 $ & $ 44.82 \pm 0.10 $ & $ 12.46 \pm 0.10 $ & $ 3.13 \pm 0.09 $ &  3C190 & $ 16759$ \\ 
J0802+614 & $ 120.74 \pm 0.17 $ & $ 61.48 \pm 0.08 $ & $ 42.35 \pm 0.10 $ & $ 38.93 \pm 0.14 $ & $ 6.29 \pm 0.11 $ & $ 1.65 \pm 0.09 $ &  - & $ 7457$ \\ 
J0806+103 & $ 121.53 \pm 0.05 $ & $ 10.31 \pm 0.07 $ & $ 57.86 \pm 0.15 $ & $ 50.79 \pm 0.15 $ & $ 13.76 \pm 0.14 $ & $ 2.37 \pm 0.08 $ &  3C191 & $ 16142$ \\ 
J0807+241 & $ 121.79 \pm 0.04 $ & $ 24.14 \pm 0.09 $ & $ 65.78 \pm 0.06 $ & $ 66.86 \pm 1.78 $ & $ 9.36 \pm 0.22 $ & $ -0.16 \pm 0.25 $ &  3C192 & $ 41548$ \\ 
J0811+423 & $ 122.91 \pm 0.08 $ & $ 42.35 \pm 0.07 $ & $ 46.74 \pm 0.07 $ & $ 49.85 \pm 0.23 $ & $ 6.30 \pm 0.13 $ & $ -0.94 \pm 0.08 $ &  - & $ 14116$ \\ 
J0815+481 & $ 123.85 \pm 0.04 $ & $ 48.19 \pm 0.08 $ & $ 157.30 \pm 0.20 $ & $ 140.87 \pm 0.31 $ & $ 30.19 \pm 0.23 $ & $ 1.61 \pm 0.04 $ &  3C196 & $ 60025$ \\ 
J0823+056 & $ 125.96 \pm 0.05 $ & $ 5.70 \pm 0.11 $ & $ 65.44 \pm 0.16 $ & $ 59.29 \pm 0.17 $ & $ 11.86 \pm 0.15 $ & $ 2.28 \pm 0.09 $ &  3C198 & $ 10087$ \\ 
J0829+292 & $ 127.30 \pm 0.05 $ & $ 29.27 \pm 0.07 $ & $ 39.14 \pm 0.08 $ & $ 35.66 \pm 0.10 $ & $ 6.52 \pm 0.08 $ & $ 1.98 \pm 0.08 $ &  3C200 & $ 11094$ \\ 
J0839+651 & $ 129.99 \pm 0.14 $ & $ 65.17 \pm 0.08 $ & $ 41.06 \pm 0.04 $ & $ 37.68 \pm 0.08 $ & $ 6.32 \pm 0.05 $ & $ 1.25 \pm 0.04 $ &  3C204 & $ 41604$ \\ 
J0841+578 & $ 130.39 \pm 0.10 $ & $ 57.89 \pm 0.11 $ & $ 43.48 \pm 0.07 $ & $ 37.62 \pm 0.10 $ & $ 10.77 \pm 0.08 $ & $ 1.64 \pm 0.04 $ &  3C205 & $ 49824$ \\ 
J0852+784 & $ 133.11 \pm 0.41 $ & $ 78.44 \pm 0.10 $ & $ 47.80 \pm 0.06 $ & $ 43.96 \pm 0.10 $ & $ 7.08 \pm 0.07 $ & $ 1.46 \pm 0.05 $ &  - & $ 29914$ \\ 
J0855+139 & $ 133.77 \pm 0.05 $ & $ 13.90 \pm 0.08 $ & $ 64.32 \pm 0.13 $ & $ 56.72 \pm 0.17 $ & $ 14.08 \pm 0.14 $ & $ 1.82 \pm 0.06 $ &  3C208 & $ 22018$ \\ 
J0859+277 & $ 134.97 \pm 0.05 $ & $ 27.80 \pm 0.07 $ & $ 36.80 \pm 0.14 $ & $ 33.13 \pm 0.14 $ & $ 7.18 \pm 0.13 $ & $ 2.40 \pm 0.15 $ &  3C210 & $ 4901$ \\ 
J0910+428 & $ 137.74 \pm 0.09 $ & $ 42.82 \pm 0.11 $ & $ 62.86 \pm 0.14 $ & $ 52.63 \pm 0.12 $ & $ 20.55 \pm 0.11 $ & $ 2.65 \pm 0.05 $ &  3C216 & $ 52948$ \\ 
J0912+377 & $ 138.09 \pm 0.08 $ & $ 37.79 \pm 0.09 $ & $ 63.12 \pm 0.12 $ & $ 54.60 \pm 0.08 $ & $ 21.17 \pm 0.09 $ & $ 4.39 \pm 0.08 $ &  3C217 & $ 55956$ \\ 
J0918+585 & $ 139.52 \pm 0.14 $ & $ 58.55 \pm 0.07 $ & $ 36.62 \pm 0.11 $ & $ 35.07 \pm 1.28 $ & $ 5.57 \pm 0.30 $ & $ 0.43 \pm 0.34 $ &  - & $ 6146$ \\ 
J0922+456 & $ 140.74 \pm 0.05 $ & $ 45.60 \pm 0.10 $ & $ 114.71 \pm 0.12 $ & $ 121.98 \pm 0.87 $ & $ 19.24 \pm 0.34 $ & $ -0.57 \pm 0.07 $ &  3C219 & $ 62283$ \\ 
J0936+790 & $ 144.24 \pm 0.39 $ & $ 79.03 \pm 0.10 $ & $ 57.90 \pm 0.05 $ & $ 52.45 \pm 0.07 $ & $ 10.01 \pm 0.05 $ & $ 1.57 \pm 0.03 $ &  - & $ 76940$ \\ 
J0941+358 & $ 145.38 \pm 0.06 $ & $ 35.84 \pm 0.08 $ & $ 40.38 \pm 0.09 $ & $ 35.13 \pm 0.05 $ & $ 14.58 \pm 0.07 $ & $ 5.38 \pm 0.10 $ &  - & $ 41495$ \\ 
J0943+137 & $ 145.95 \pm 0.05 $ & $ 13.79 \pm 0.09 $ & $ 55.56 \pm 0.15 $ & $ 47.24 \pm 0.13 $ & $ 16.77 \pm 0.12 $ & $ 2.68 \pm 0.07 $ &  3C225 & $ 23350$ \\ 
J0943+831 & $ 145.97 \pm 0.59 $ & $ 83.18 \pm 0.10 $ & $ 59.12 \pm 0.05 $ & $ 53.30 \pm 0.09 $ & $ 10.70 \pm 0.06 $ & $ 1.47 \pm 0.03 $ &  - & $ 88487$ \\ 
J0945+097 & $ 146.35 \pm 0.05 $ & $ 9.71 \pm 0.06 $ & $ 57.78 \pm 0.17 $ & $ 50.83 \pm 0.15 $ & $ 14.01 \pm 0.14 $ & $ 2.68 \pm 0.09 $ &  3C226 & $ 13561$ \\ 
J0949+073 & $ 147.37 \pm 0.04 $ & $ 7.36 \pm 0.07 $ & $ 113.72 \pm 0.12 $ & $ 106.25 \pm 0.17 $ & $ 13.73 \pm 0.14 $ & $ 1.65 \pm 0.05 $ &  3C227 & $ 18700$ \\ 
J0951+143 & $ 147.95 \pm 0.06 $ & $ 14.34 \pm 0.09 $ & $ 50.78 \pm 0.10 $ & $ 45.14 \pm 0.11 $ & $ 10.71 \pm 0.09 $ & $ 2.14 \pm 0.06 $ &  3C228 & $ 22500$ \\ 
J0952+731 & $ 148.11 \pm 0.21 $ & $ 73.18 \pm 0.10 $ & $ 49.90 \pm 0.04 $ & $ 47.81 \pm 0.41 $ & $ 6.19 \pm 0.13 $ & $ 0.49 \pm 0.10 $ &  - & $ 64167$ \\ 
J0958+694 & $ 149.70 \pm 0.17 $ & $ 69.48 \pm 0.09 $ & $ 41.65 \pm 0.07 $ & $ 37.83 \pm 0.15 $ & $ 7.14 \pm 0.10 $ & $ 1.27 \pm 0.06 $ &  3C231 & $ 24283$ \\ 
J1003+287 & $ 150.85 \pm 0.05 $ & $ 28.72 \pm 0.08 $ & $ 90.06 \pm 0.15 $ & $ 76.73 \pm 0.11 $ & $ 31.20 \pm 0.14 $ & $ 3.90 \pm 0.06 $ &  3C234 & $ 47323$ \\ 
J1006+225 & $ 151.53 \pm 0.23 $ & $ 22.54 \pm 0.09 $ & $ 46.39 \pm 0.20 $ & $ 40.41 \pm 0.28 $ & $ 11.03 \pm 0.21 $ & $ 1.71 \pm 0.11 $ &  - & $ 13088$ \\ 
J1007+347 & $ 151.94 \pm 0.08 $ & $ 34.78 \pm 0.09 $ & $ 39.54 \pm 0.11 $ & $ 34.27 \pm 0.06 $ & $ 16.44 \pm 0.08 $ & $ 6.49 \pm 0.14 $ &  - & $ 29749$ \\ 
J1011+068 & $ 152.77 \pm 0.24 $ & $ 6.88 \pm 0.38 $ & $ 61.42 \pm 0.22 $ & $ 52.52 \pm 0.23 $ & $ 17.18 \pm 0.21 $ & $ 2.31 \pm 0.09 $ &  3C237 & $ 10461$ \\ 
J1013+463 & $ 153.39 \pm 0.09 $ & $ 46.34 \pm 0.09 $ & $ 46.28 \pm 0.08 $ & $ 39.62 \pm 0.07 $ & $ 13.44 \pm 0.07 $ & $ 2.70 \pm 0.05 $ &  - & $ 49609$ \\ 
J1016+745 & $ 154.15 \pm 0.30 $ & $ 74.56 \pm 0.11 $ & $ 42.11 \pm 0.14 $ & $ 37.26 \pm 0.13 $ & $ 9.64 \pm 0.12 $ & $ 2.56 \pm 0.12 $ &  - & $ 15676$ \\ 
J1020+808 & $ 155.09 \pm 0.65 $ & $ 80.83 \pm 0.14 $ & $ 52.86 \pm 0.11 $ & $ 46.47 \pm 0.08 $ & $ 14.22 \pm 0.09 $ & $ 3.48 \pm 0.08 $ &  - & $ 28682$ \\ 
J1027+064 & $ 156.89 \pm 0.04 $ & $ 6.49 \pm 0.09 $ & $ 54.13 \pm 0.13 $ & $ 49.08 \pm 0.17 $ & $ 9.39 \pm 0.14 $ & $ 1.88 \pm 0.09 $ &  3C243 & $ 8357$ \\ 
J1035+581 & $ 158.84 \pm 0.10 $ & $ 58.19 \pm 0.10 $ & $ 58.04 \pm 0.09 $ & $ 50.83 \pm 0.24 $ & $ 13.91 \pm 0.14 $ & $ 1.08 \pm 0.04 $ &  - & $ 73615$ \\ 
J1044+118 & $ 161.09 \pm 0.05 $ & $ 11.86 \pm 0.08 $ & $ 42.83 \pm 0.14 $ & $ 38.72 \pm 0.12 $ & $ 8.52 \pm 0.12 $ & $ 2.91 \pm 0.14 $ &  3C245 & $ 5979$ \\ 
J1108+768 & $ 167.23 \pm 0.31 $ & $ 76.86 \pm 0.10 $ & $ 51.76 \pm 0.05 $ & $ 46.84 \pm 0.09 $ & $ 9.14 \pm 0.06 $ & $ 1.31 \pm 0.03 $ &  - & $ 62816$ \\ 
J1110+249 & $ 167.61 \pm 0.07 $ & $ 24.96 \pm 0.08 $ & $ 48.67 \pm 0.08 $ & $ 42.94 \pm 0.08 $ & $ 11.28 \pm 0.07 $ & $ 2.50 \pm 0.05 $ &  3C250 & $ 34110$ \\ 
J1113+355 & $ 168.29 \pm 0.07 $ & $ 35.59 \pm 0.09 $ & $ 41.12 \pm 0.12 $ & $ 35.15 \pm 0.06 $ & $ 21.53 \pm 0.09 $ & $ 8.05 \pm 0.16 $ &  3C252 & $ 41489$ \\ 
J1115+405 & $ 168.93 \pm 0.07 $ & $ 40.58 \pm 0.09 $ & $ 61.78 \pm 0.15 $ & $ 50.50 \pm 0.09 $ & $ 31.24 \pm 0.12 $ & $ 5.34 \pm 0.08 $ &  3C254 & $ 61784$ \\ 
J1141+657 & $ 175.32 \pm 0.18 $ & $ 65.71 \pm 0.09 $ & $ 50.40 \pm 0.07 $ & $ 43.84 \pm 0.12 $ & $ 12.07 \pm 0.08 $ & $ 1.46 \pm 0.03 $ &  3C263 & $ 73340$ \\ 
J1144+218 & $ 176.09 \pm 0.07 $ & $ 21.89 \pm 0.09 $ & $ 50.02 \pm 0.16 $ & $ 42.33 \pm 0.10 $ & $ 20.27 \pm 0.13 $ & $ 4.91 \pm 0.12 $ &  - & $ 20454$ \\ 
J1146+195 & $ 176.60 \pm 0.05 $ & $ 19.59 \pm 0.09 $ & $ 80.30 \pm 0.13 $ & $ 69.63 \pm 0.15 $ & $ 20.26 \pm 0.13 $ & $ 2.13 \pm 0.04 $ &  3C264 & $ 39290$ \\ 
J1146+315 & $ 176.75 \pm 0.05 $ & $ 31.54 \pm 0.09 $ & $ 60.55 \pm 0.11 $ & $ 52.09 \pm 0.16 $ & $ 15.60 \pm 0.12 $ & $ 1.70 \pm 0.05 $ &  3C265 & $ 49055$ \\ 
J1147+496 & $ 176.84 \pm 0.07 $ & $ 49.65 \pm 0.06 $ & $ 36.86 \pm 0.15 $ & $ 31.98 \pm 0.12 $ & $ 10.44 \pm 0.12 $ & $ 3.18 \pm 0.13 $ &  - & $ 8194$ \\ 
J1151+127 & $ 177.84 \pm 0.05 $ & $ 12.74 \pm 0.09 $ & $ 49.69 \pm 0.15 $ & $ 44.88 \pm 0.19 $ & $ 8.89 \pm 0.16 $ & $ 1.80 \pm 0.10 $ &  3C267 & $ 5008$ \\ 
J1156+547 & $ 179.17 \pm 0.10 $ & $ 54.76 \pm 0.08 $ & $ 36.21 \pm 0.08 $ & $ 31.97 \pm 0.06 $ & $ 9.59 \pm 0.07 $ & $ 3.61 \pm 0.09 $ &  - & $ 19741$ \\ 
J1200+729 & $ 180.25 \pm 0.24 $ & $ 72.93 \pm 0.11 $ & $ 54.85 \pm 0.09 $ & $ 47.09 \pm 0.14 $ & $ 14.26 \pm 0.10 $ & $ 1.59 \pm 0.04 $ &  - & $ 91796$ \\ 
J1210+436 & $ 182.68 \pm 0.07 $ & $ 43.62 \pm 0.06 $ & $ 33.98 \pm 0.10 $ & $ 30.00 \pm 0.09 $ & $ 8.13 \pm 0.08 $ & $ 2.80 \pm 0.10 $ &  - & $ 11854$ \\ 
J1217+534 & $ 184.36 \pm 0.09 $ & $ 53.46 \pm 0.05 $ & $ 34.14 \pm 0.12 $ & $ 31.09 \pm 0.13 $ & $ 5.90 \pm 0.12 $ & $ 2.31 \pm 0.15 $ &  - & $ 4537$ \\ 
J1220+057 & $ 185.17 \pm 0.06 $ & $ 5.77 \pm 0.09 $ & $ 123.19 \pm 0.14 $ & $ 113.56 \pm 0.18 $ & $ 17.76 \pm 0.15 $ & $ 1.72 \pm 0.05 $ &  3C270 & $ 20649$ \\ 
J1222+336 & $ 185.51 \pm 0.05 $ & $ 33.62 \pm 0.06 $ & $ 37.49 \pm 0.11 $ & $ 32.39 \pm 0.09 $ & $ 11.15 \pm 0.09 $ & $ 3.35 \pm 0.10 $ &  - & $ 15260$ \\ 
J1236+212 & $ 189.21 \pm 0.06 $ & $ 21.23 \pm 0.08 $ & $ 46.24 \pm 0.10 $ & $ 40.99 \pm 0.11 $ & $ 9.93 \pm 0.09 $ & $ 2.10 \pm 0.06 $ &  - & $ 18931$ \\ 
J1245+163 & $ 191.30 \pm 0.06 $ & $ 16.33 \pm 0.09 $ & $ 54.99 \pm 0.23 $ & $ 48.99 \pm 0.34 $ & $ 11.05 \pm 0.25 $ & $ 1.60 \pm 0.12 $ &  - & $ 4432$ \\ 
J1255+156 & $ 193.76 \pm 0.07 $ & $ 15.64 \pm 0.08 $ & $ 51.29 \pm 0.18 $ & $ 46.39 \pm 0.24 $ & $ 9.05 \pm 0.19 $ & $ 1.75 \pm 0.12 $ &  - & $ 4450$ \\ 
J1257+473 & $ 194.42 \pm 0.09 $ & $ 47.31 \pm 0.10 $ & $ 55.22 \pm 0.11 $ & $ 45.59 \pm 0.08 $ & $ 21.78 \pm 0.08 $ & $ 3.61 \pm 0.05 $ &  3C280 & $ 67093$ \\ 
J1259+278 & $ 194.97 \pm 0.13 $ & $ 27.82 \pm 0.08 $ & $ 58.63 \pm 0.06 $ & $ 59.12 \pm 1.54 $ & $ 10.83 \pm 0.12 $ & $ -0.06 \pm 0.18 $ &  - & $ 34053$ \\ 
J1312+273 & $ 198.04 \pm 0.07 $ & $ 27.40 \pm 0.07 $ & $ 43.60 \pm 0.11 $ & $ 38.86 \pm 0.12 $ & $ 8.96 \pm 0.11 $ & $ 2.09 \pm 0.08 $ &  3C284 & $ 11552$ \\ 
J1332+305 & $ 203.01 \pm 0.09 $ & $ 30.52 \pm 0.09 $ & $ 47.69 \pm 0.12 $ & $ 41.59 \pm 0.10 $ & $ 12.27 \pm 0.10 $ & $ 2.67 \pm 0.07 $ &  3C286 & $ 21608$ \\ 
J1336+409 & $ 204.08 \pm 0.10 $ & $ 40.93 \pm 0.06 $ & $ 34.73 \pm 0.11 $ & $ 31.98 \pm 0.15 $ & $ 5.09 \pm 0.12 $ & $ 1.75 \pm 0.13 $ &  - & $ 3943$ \\ 
J1340+387 & $ 205.07 \pm 0.07 $ & $ 38.79 \pm 0.08 $ & $ 48.08 \pm 0.07 $ & $ 46.33 \pm 1.39 $ & $ 8.54 \pm 0.25 $ & $ 0.28 \pm 0.22 $ &  3C288 & $ 31456$ \\ 
J1346+496 & $ 206.71 \pm 0.09 $ & $ 49.65 \pm 0.06 $ & $ 40.64 \pm 0.13 $ & $ 37.04 \pm 0.23 $ & $ 6.69 \pm 0.16 $ & $ 1.35 \pm 0.11 $ &  - & $ 5813$ \\ 
J1351+643 & $ 207.78 \pm 0.15 $ & $ 64.36 \pm 0.09 $ & $ 47.26 \pm 0.07 $ & $ 42.09 \pm 0.15 $ & $ 9.85 \pm 0.10 $ & $ 1.14 \pm 0.04 $ &  3C292 & $ 45160$ \\ 
J1354+315 & $ 208.66 \pm 0.13 $ & $ 31.54 \pm 0.08 $ & $ 50.76 \pm 0.09 $ & $ 47.04 \pm 0.36 $ & $ 8.03 \pm 0.18 $ & $ 0.81 \pm 0.09 $ &  3C293 & $ 16714$ \\ 
J1407+341 & $ 211.89 \pm 0.08 $ & $ 34.13 \pm 0.07 $ & $ 48.77 \pm 0.14 $ & $ 41.17 \pm 0.15 $ & $ 14.38 \pm 0.12 $ & $ 2.11 \pm 0.06 $ &  3C294 & $ 31019$ \\ 
J1410+313 & $ 212.67 \pm 0.08 $ & $ 31.33 \pm 0.08 $ & $ 46.28 \pm 0.22 $ & $ 48.81 \pm 1.62 $ & $ 7.01 \pm 0.50 $ & $ -0.62 \pm 0.38 $ &  - & $ 4231$ \\ 
J1412+521 & $ 213.15 \pm 0.05 $ & $ 52.17 \pm 0.08 $ & $ 112.55 \pm 0.18 $ & $ 95.06 \pm 0.12 $ & $ 41.24 \pm 0.14 $ & $ 3.96 \pm 0.05 $ &  3C295 & $ 77864$ \\ 
J1420+064 & $ 215.11 \pm 0.04 $ & $ 6.48 \pm 0.07 $ & $ 86.31 \pm 0.39 $ & $ 67.76 \pm 0.27 $ & $ 43.04 \pm 0.29 $ & $ 3.82 \pm 0.10 $ &  3C298 & $ 21751$ \\ 
J1422+416 & $ 215.61 \pm 0.08 $ & $ 41.69 \pm 0.07 $ & $ 41.98 \pm 0.12 $ & $ 36.64 \pm 0.10 $ & $ 11.29 \pm 0.10 $ & $ 3.08 \pm 0.09 $ &  3C299 & $ 12810$ \\ 
J1424+195 & $ 216.18 \pm 0.06 $ & $ 19.52 \pm 0.11 $ & $ 60.66 \pm 0.18 $ & $ 51.91 \pm 0.09 $ & $ 27.69 \pm 0.12 $ & $ 6.63 \pm 0.14 $ &  - & $ 35518$ \\ 
J1427+376 & $ 216.83 \pm 0.07 $ & $ 37.65 \pm 0.06 $ & $ 38.91 \pm 0.16 $ & $ 35.79 \pm 0.67 $ & $ 6.48 \pm 0.29 $ & $ 0.91 \pm 0.20 $ &  - & $ 4503$ \\ 
J1444+519 & $ 221.04 \pm 0.09 $ & $ 51.98 \pm 0.07 $ & $ 39.45 \pm 0.16 $ & $ 35.18 \pm 0.13 $ & $ 8.99 \pm 0.14 $ & $ 3.03 \pm 0.15 $ &  3C303 & $ 5025$ \\ 
J1445+768 & $ 221.46 \pm 0.43 $ & $ 76.90 \pm 0.09 $ & $ 45.58 \pm 0.10 $ & $ 39.13 \pm 0.11 $ & $ 12.32 \pm 0.09 $ & $ 2.19 \pm 0.06 $ &  - & $ 37122$ \\ 
J1449+631 & $ 222.47 \pm 0.20 $ & $ 63.18 \pm 0.08 $ & $ 43.26 \pm 0.08 $ & $ 37.74 \pm 0.06 $ & $ 12.35 \pm 0.06 $ & $ 3.53 \pm 0.06 $ &  3C305 & $ 33402$ \\ 
J1501+714 & $ 225.29 \pm 0.24 $ & $ 71.48 \pm 0.20 $ & $ 44.77 \pm 0.13 $ & $ 38.12 \pm 0.07 $ & $ 21.64 \pm 0.10 $ & $ 6.92 \pm 0.14 $ &  - & $ 32787$ \\ 
J1506+259 & $ 226.55 \pm 0.03 $ & $ 25.97 \pm 0.07 $ & $ 127.36 \pm 0.13 $ & $ 112.69 \pm 0.11 $ & $ 31.71 \pm 0.12 $ & $ 3.25 \pm 0.04 $ &  3C310 & $ 57615$ \\ 
J1518+070 & $ 229.53 \pm 0.03 $ & $ 7.01 \pm 0.06 $ & $ 114.72 \pm 0.12 $ & $ 117.79 \pm 2.38 $ & $ 13.00 \pm 0.41 $ & $ -0.33 \pm 0.25 $ &  3C317 & $ 31047$ \\ 
J1523+076 & $ 230.82 \pm 0.04 $ & $ 7.66 \pm 0.08 $ & $ 71.86 \pm 0.13 $ & $ 66.71 \pm 0.23 $ & $ 9.52 \pm 0.16 $ & $ 1.37 \pm 0.08 $ &  - & $ 8780$ \\ 
J1525+543 & $ 231.31 \pm 0.09 $ & $ 54.38 \pm 0.08 $ & $ 46.64 \pm 0.04 $ & $ 47.26 \pm 1.18 $ & $ 7.23 \pm 0.13 $ & $ -0.11 \pm 0.21 $ &  3C319 & $ 42515$ \\ 
J1533+239 & $ 233.26 \pm 0.06 $ & $ 23.96 \pm 0.08 $ & $ 57.89 \pm 0.13 $ & $ 49.36 \pm 0.10 $ & $ 19.71 \pm 0.10 $ & $ 3.79 \pm 0.07 $ &  3C321 & $ 40369$ \\ 
J1551+626 & $ 237.76 \pm 0.12 $ & $ 62.65 \pm 0.07 $ & $ 42.27 \pm 0.13 $ & $ 37.38 \pm 0.10 $ & $ 10.77 \pm 0.10 $ & $ 3.39 \pm 0.12 $ &  3C325 & $ 12332$ \\ 
J1611+658 & $ 242.77 \pm 0.14 $ & $ 65.88 \pm 0.10 $ & $ 50.32 \pm 0.10 $ & $ 42.51 \pm 0.05 $ & $ 24.82 \pm 0.08 $ & $ 6.67 \pm 0.09 $ &  3C330 & $ 69223$ \\ 
J1629+395 & $ 247.44 \pm 0.04 $ & $ 39.51 \pm 0.07 $ & $ 120.52 \pm 0.11 $ & $ 106.56 \pm 0.09 $ & $ 29.71 \pm 0.11 $ & $ 3.13 \pm 0.03 $ &  3C338 & $ 62467$ \\ 
J1629+823 & $ 247.49 \pm 0.56 $ & $ 82.32 \pm 0.06 $ & $ 46.36 \pm 0.08 $ & $ 43.33 \pm 0.12 $ & $ 5.57 \pm 0.09 $ & $ 1.55 \pm 0.08 $ &  - & $ 8240$ \\ 
J1630+442 & $ 247.62 \pm 0.09 $ & $ 44.30 \pm 0.10 $ & $ 48.80 \pm 0.19 $ & $ 42.93 \pm 0.18 $ & $ 11.72 \pm 0.16 $ & $ 2.61 \pm 0.12 $ &  3C337 & $ 8597$ \\ 
J1652+050 & $ 253.12 \pm 0.02 $ & $ 5.01 \pm 0.05 $ & $ 873.07 \pm 0.37 $ & $ 863.96 \pm 7.50 $ & $ 39.98 \pm 1.35 $ & $ 0.32 \pm 0.26 $ &  3C348 & $ 23233$ \\ 
J1711+460 & $ 257.94 \pm 0.11 $ & $ 46.01 \pm 0.08 $ & $ 44.38 \pm 0.13 $ & $ 40.47 \pm 0.10 $ & $ 8.84 \pm 0.11 $ & $ 3.60 \pm 0.15 $ &  3C352 & $ 7215$ \\ 
J1725+508 & $ 261.27 \pm 0.14 $ & $ 50.84 \pm 0.07 $ & $ 41.84 \pm 0.20 $ & $ 36.15 \pm 0.12 $ & $ 14.58 \pm 0.14 $ & $ 4.66 \pm 0.19 $ &  3C356 & $ 7369$ \\ 
J1746+802 & $ 266.60 \pm 0.58 $ & $ 80.27 \pm 0.09 $ & $ 60.42 \pm 0.16 $ & $ 54.18 \pm 0.31 $ & $ 11.60 \pm 0.21 $ & $ 1.32 \pm 0.08 $ &  - & $ 14183$ \\ 
J1830+487 & $ 277.65 \pm 0.04 $ & $ 48.74 \pm 0.08 $ & $ 136.75 \pm 0.09 $ & $ 127.96 \pm 0.19 $ & $ 16.70 \pm 0.12 $ & $ 1.15 \pm 0.03 $ &  3C380 & $ 62202$ \\ 
J1836+326 & $ 279.13 \pm 0.07 $ & $ 32.69 \pm 0.08 $ & $ 56.65 \pm 0.22 $ & $ 50.16 \pm 0.15 $ & $ 15.77 \pm 0.17 $ & $ 4.20 \pm 0.17 $ &  - & $ 7804$ \\ 
J1840+797 & $ 280.22 \pm 0.55 $ & $ 79.78 \pm 0.07 $ & $ 117.25 \pm 0.14 $ & $ 104.40 \pm 0.14 $ & $ 24.73 \pm 0.12 $ & $ 2.27 \pm 0.04 $ &  - & $ 121769$ \\ 
J1844+455 & $ 281.22 \pm 0.07 $ & $ 45.56 \pm 0.06 $ & $ 54.96 \pm 0.07 $ & $ 49.80 \pm 0.07 $ & $ 10.00 \pm 0.07 $ & $ 2.35 \pm 0.05 $ &  3C388 & $ 24996$ \\ 
J1941+607 & $ 295.42 \pm 0.14 $ & $ 60.71 \pm 0.09 $ & $ 44.88 \pm 0.08 $ & $ 40.55 \pm 0.11 $ & $ 7.97 \pm 0.09 $ & $ 1.70 \pm 0.06 $ &  3C401 & $ 17535$ \\ 
J1955+328 & $ 298.93 \pm 0.08 $ & $ 32.88 \pm 0.10 $ & $ 114.73 \pm 0.24 $ & $ 115.93 \pm 3.51 $ & $ 12.29 \pm 0.54 $ & $ -0.15 \pm 0.40 $ &  CTB 80 & $ 5553$ \\ 
J2015+236 & $ 303.94 \pm 0.03 $ & $ 23.61 \pm 0.07 $ & $ 206.18 \pm 0.20 $ & $ 214.69 \pm 2.21 $ & $ 21.56 \pm 0.54 $ & $ -0.62 \pm 0.15 $ &  3C409 & $ 30016$ \\ 
J2020+455 & $ 305.17 \pm 0.04 $ & $ 45.56 \pm 0.08 $ & $ 288.53 \pm 0.13 $ & $ 277.60 \pm 0.30 $ & $ 21.02 \pm 0.19 $ & $ 1.09 \pm 0.04 $ &  W63 & $ 50894$ \\ 
J2021+297 & $ 305.39 \pm 0.05 $ & $ 29.73 \pm 0.06 $ & $ 103.62 \pm 0.14 $ & $ 97.94 \pm 0.43 $ & $ 11.31 \pm 0.25 $ & $ 0.98 \pm 0.09 $ &  3C410 & $ 17491$ \\ 
J2022+403 & $ 305.61 \pm 0.02 $ & $ 40.37 \pm 0.07 $ & $ 671.39 \pm 0.29 $ & $ 644.72 \pm 0.43 $ & $ 49.01 \pm 0.33 $ & $ 1.57 \pm 0.03 $ &  DR4 & $ 46342$ \\ 
J2046+506 & $ 311.61 \pm 0.05 $ & $ 50.65 \pm 0.09 $ & $ 285.26 \pm 0.25 $ & $ 266.13 \pm 0.21 $ & $ 39.82 \pm 0.21 $ & $ 2.95 \pm 0.05 $ &  HB21 & $ 52400$ \\ 
J2051+300 & $ 312.85 \pm 0.04 $ & $ 30.01 \pm 0.08 $ & $ 191.25 \pm 0.24 $ & $ 177.54 \pm 0.51 $ & $ 25.73 \pm 0.33 $ & $ 1.23 \pm 0.06 $ &  Cygnus Loop & $ 29773$ \\ 
J2103+766 & $ 315.77 \pm 0.41 $ & $ 76.63 \pm 0.09 $ & $ 68.78 \pm 0.14 $ & $ 57.08 \pm 0.11 $ & $ 24.72 \pm 0.10 $ & $ 3.07 \pm 0.05 $ &  - & $ 110618$ \\ 
J2118+608 & $ 319.75 \pm 0.09 $ & $ 60.88 \pm 0.08 $ & $ 96.04 \pm 0.07 $ & $ 89.62 \pm 0.13 $ & $ 12.08 \pm 0.09 $ & $ 1.21 \pm 0.03 $ &  3C430 & $ 56715$ \\ 
J2124+251 & $ 321.25 \pm 0.03 $ & $ 25.12 \pm 0.06 $ & $ 128.30 \pm 0.13 $ & $ 120.52 \pm 0.24 $ & $ 14.55 \pm 0.16 $ & $ 1.26 \pm 0.05 $ &  3C433 & $ 23010$ \\ 
J2145+281 & $ 326.39 \pm 0.05 $ & $ 28.16 \pm 0.09 $ & $ 59.12 \pm 0.18 $ & $ 52.60 \pm 0.11 $ & $ 16.92 \pm 0.15 $ & $ 4.77 \pm 0.15 $ &  3C436 & $ 11360$ \\ 
J2157+380 & $ 329.30 \pm 0.04 $ & $ 38.04 \pm 0.08 $ & $ 114.55 \pm 0.12 $ & $ 116.94 \pm 2.86 $ & $ 12.56 \pm 0.40 $ & $ -0.27 \pm 0.31 $ &  3C438 & $ 30219$ \\ 
J2247+397 & $ 341.79 \pm 0.03 $ & $ 39.73 \pm 0.08 $ & $ 150.89 \pm 0.15 $ & $ 161.54 \pm 0.27 $ & $ 19.69 \pm 0.19 $ & $ -1.37 \pm 0.04 $ &  3C452 & $ 26727$ \\ 
J2339+271 & $ 354.98 \pm 0.04 $ & $ 27.10 \pm 0.07 $ & $ 119.61 \pm 0.18 $ & $ 116.70 \pm 3.34 $ & $ 16.11 \pm 0.57 $ & $ 0.25 \pm 0.28 $ &  3C465 & $ 17031$ \\ 
\hline \hline 
\caption{The first 10 source, as an example, of the AARTFAAC northern hemisphere catalogue at 60MHz. The full catalogue as a machine readable csv is available online. The labels here are derived from the positions measured, as in the VLSSr, the positions are given in degrees with the uncertainty  calculated from the standard deviation of the individual position measurements. The flux density values, and uncertainties, are calculated from the mode of a skew-normal distribution fit to the population of measurements. The parameters of the skew normal are given. Common names are given  to  matching sources where possible. Lastly the number of images from which the lightcurve is generated is given.  }
\label{tab:aartfaaccat}
\end{longtable}

\end{landscape}

\section{Supplementary online material: Helmbolt spectra comparison}

Here the flux density as measured by AARTFAAC, open circles, are compared to the measurements and spectral fits by \citet{2008ApJS..174..313H}, the red dots are the measured flux densities, the dashed lines are power law fits, and the solid lines power laws with turnover at lower frequencies. Some sources have not had any spectra fit, while others are clearly bad fits. In all cases these are taken directly from \citet{2008ApJS..174..313H}, without considering the AARTFAAC flux density value, for comparison purposes. For additional comparisons, the plot of Tycho supernova remnant, top left, includes single dish measurements made by \citet{1979A&A....76..120K}, blue circles. Lastly, where source associations could be made the 8C catalogue \citep{1990MNRAS.244..233R} flux density values are plotted (green circles.) Given that the AARTFAAC catalogue was bootstrapped entirely from eight \citet{2017ApJS..230....7P} sources, proper agreement across this broad range of sources is an excellent, independent, test of the validity of the calibration method, and the flux density scaling.  

\label{apx:spectra}

\begin{figure}
\begin{center}
\includegraphics[height=0.95\textheight]{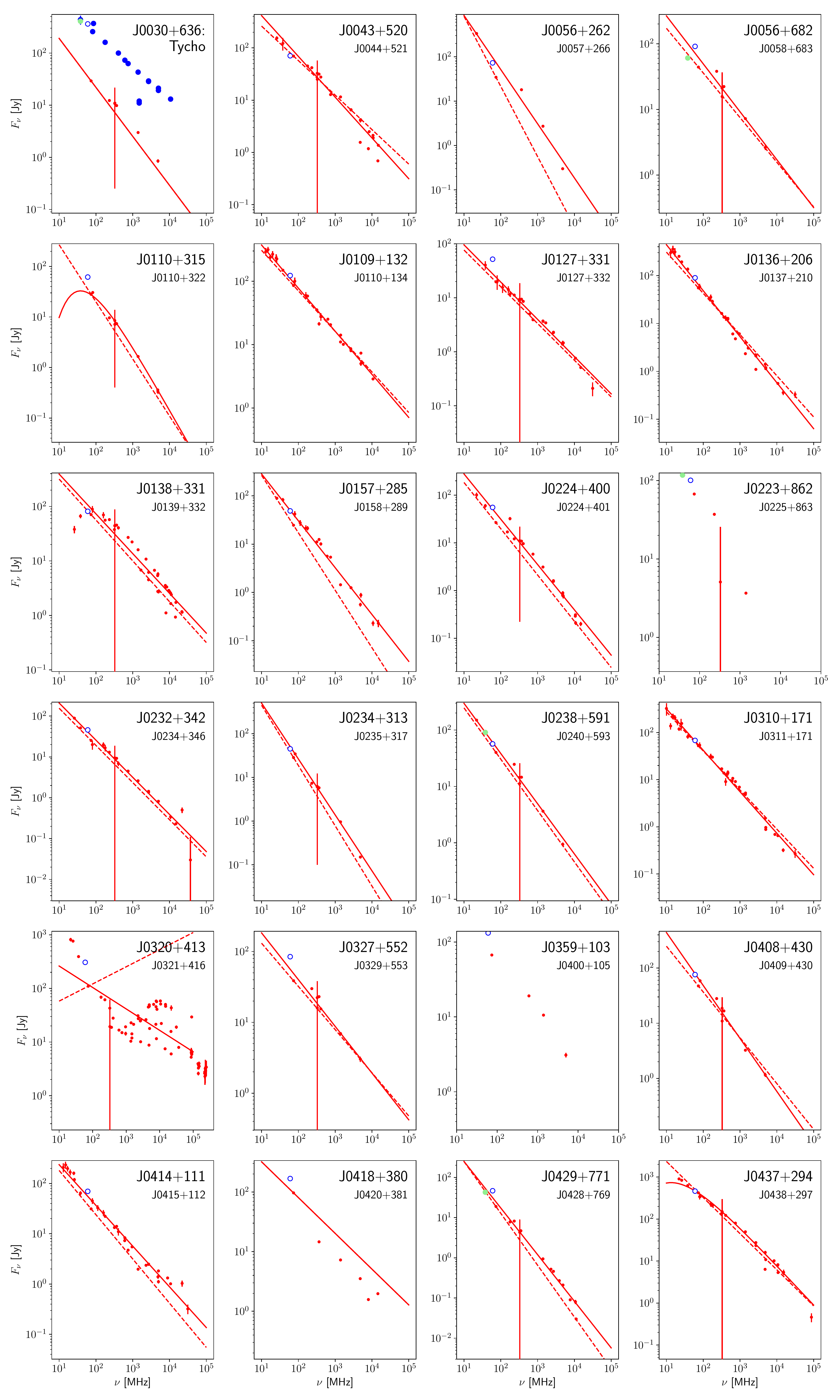}
\caption{Here the flux density as measured by AARTFAAC (open circles) are compared to the measurements and spectra fit by Helmbolt et al. (2008) (red dots and lines) measurements by Klein et al. (1979) (blue circles) and Rees et al. (1990) (green circles). The larger label refers to the designation assigned by Helmbolt et al. (2008), and the smaller label below is the label in the AARTFAAC catalogue.} 
\label{fig:helm_compare_100}
\end{center}
\end{figure}

\begin{figure}
\begin{center}
\includegraphics[height=0.95\textheight]{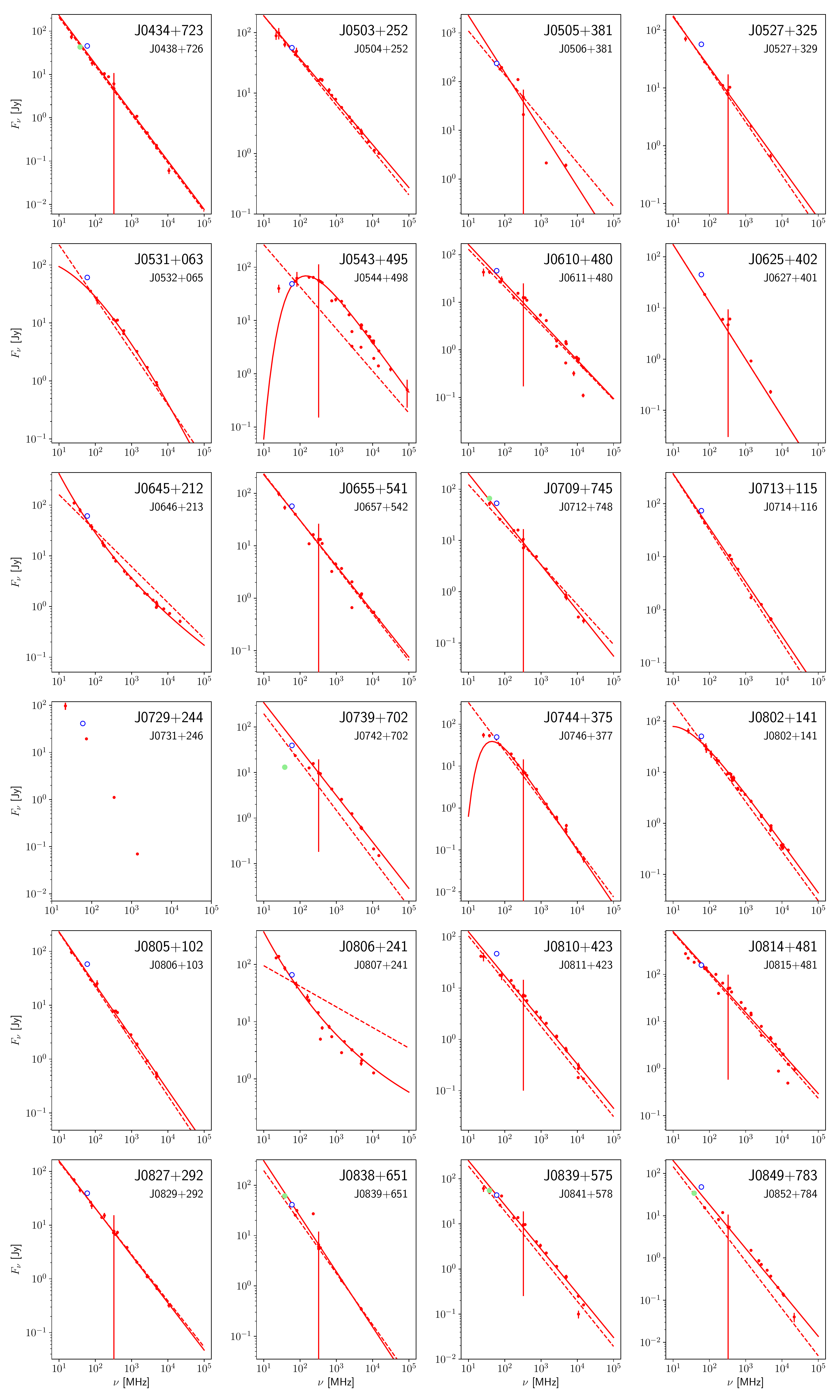}
\caption{Here the flux density as measured by AARTFAAC (open circles) are compared to the measurements and spectra fit by Helmbolt et al. (2008) (red dots and lines) measurements by Klein et al. (1979) (blue circles) and Rees et al. (1990) (green circles). The larger label refers to the designation assigned by Helmbolt et al. (2008), and the smaller label below is the label in the AARTFAAC catalogue.} 
\label{fig:helm_compare_101}
\end{center}
\end{figure}

\begin{figure}
\begin{center}
\includegraphics[height=0.95\textheight]{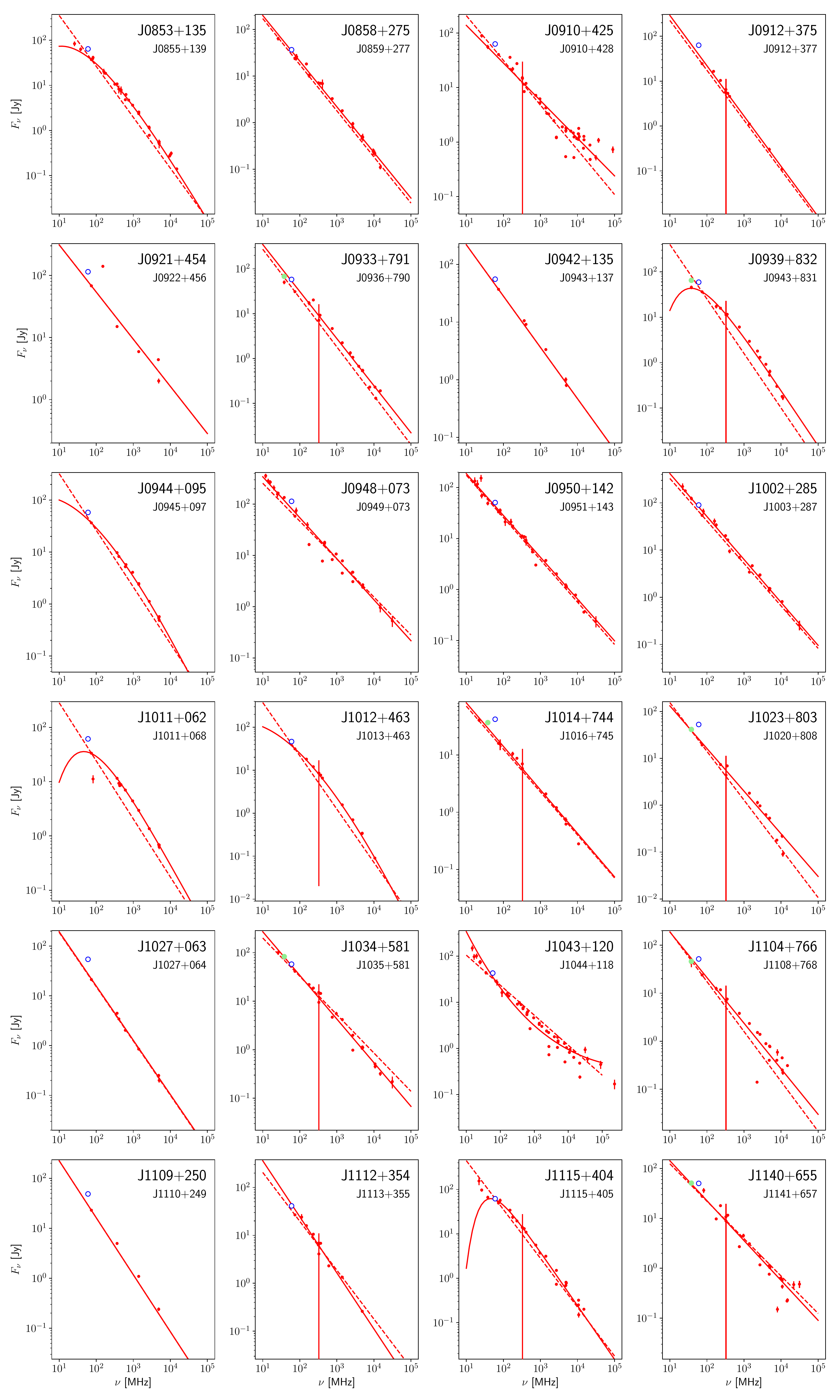}
\caption{Here the flux density as measured by AARTFAAC (open circles) are compared to the measurements and spectra fit by Helmbolt et al. (2008) (red dots and lines) measurements by Klein et al. (1979) (blue circles) and Rees et al. (1990) (green circles). The larger label refers to the designation assigned by Helmbolt et al. (2008), and the smaller label below is the label in the AARTFAAC catalogue.} 
\label{fig:helm_compare_102}
\end{center}
\end{figure}

\begin{figure}
\begin{center}
\includegraphics[height=0.95\textheight]{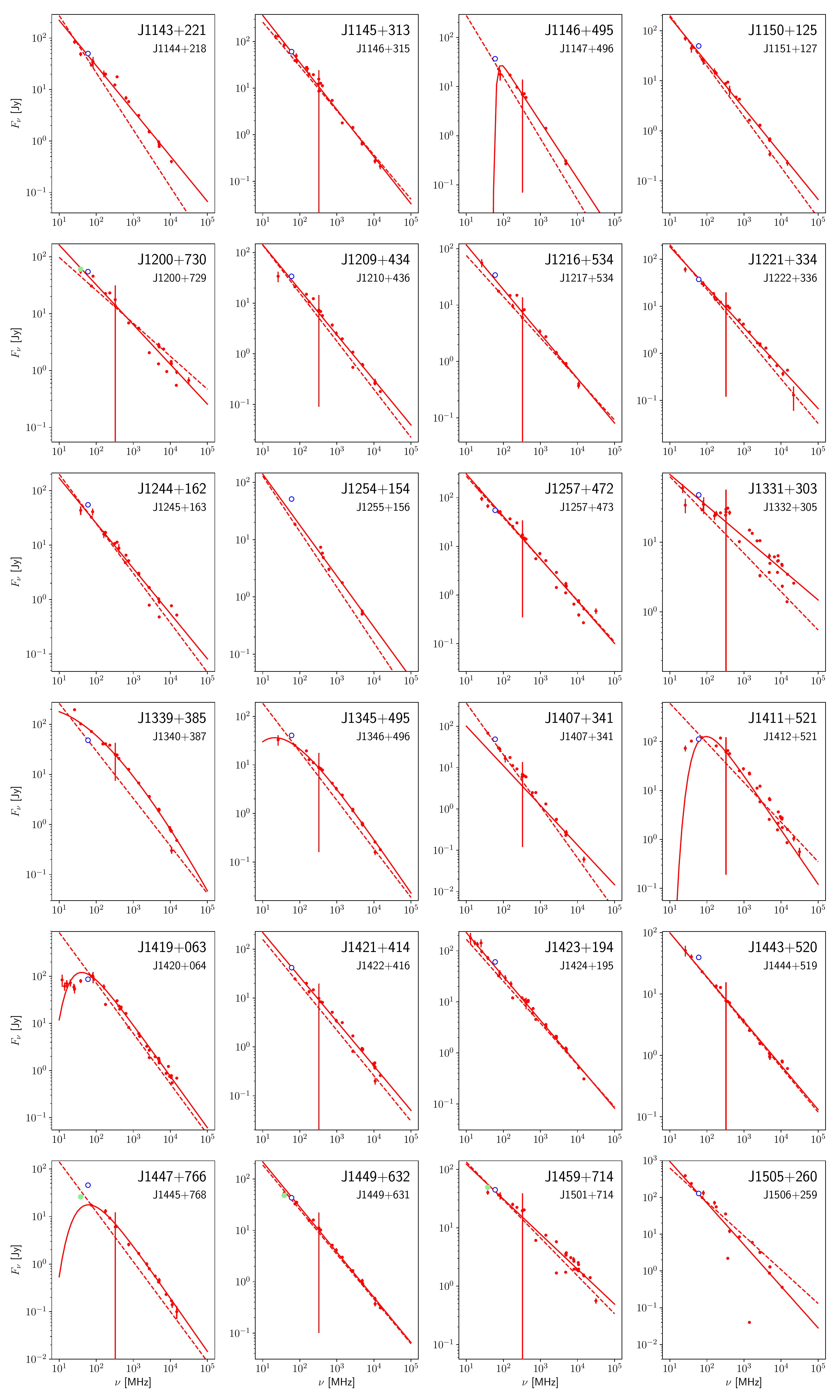}
\caption{Here the flux density as measured by AARTFAAC (open circles) are compared to the measurements and spectra fit by Helmbolt et al. (2008) (red dots and lines) measurements by Klein et al. (1979) (blue circles) and Rees et al. (1990) (green circles). The larger label refers to the designation assigned by Helmbolt et al. (2008), and the smaller label below is the label in the AARTFAAC catalogue.} 
\label{fig:helm_compare_103}
\end{center}
\end{figure}

\begin{figure}
\begin{center}
\includegraphics[height=0.95\textheight]{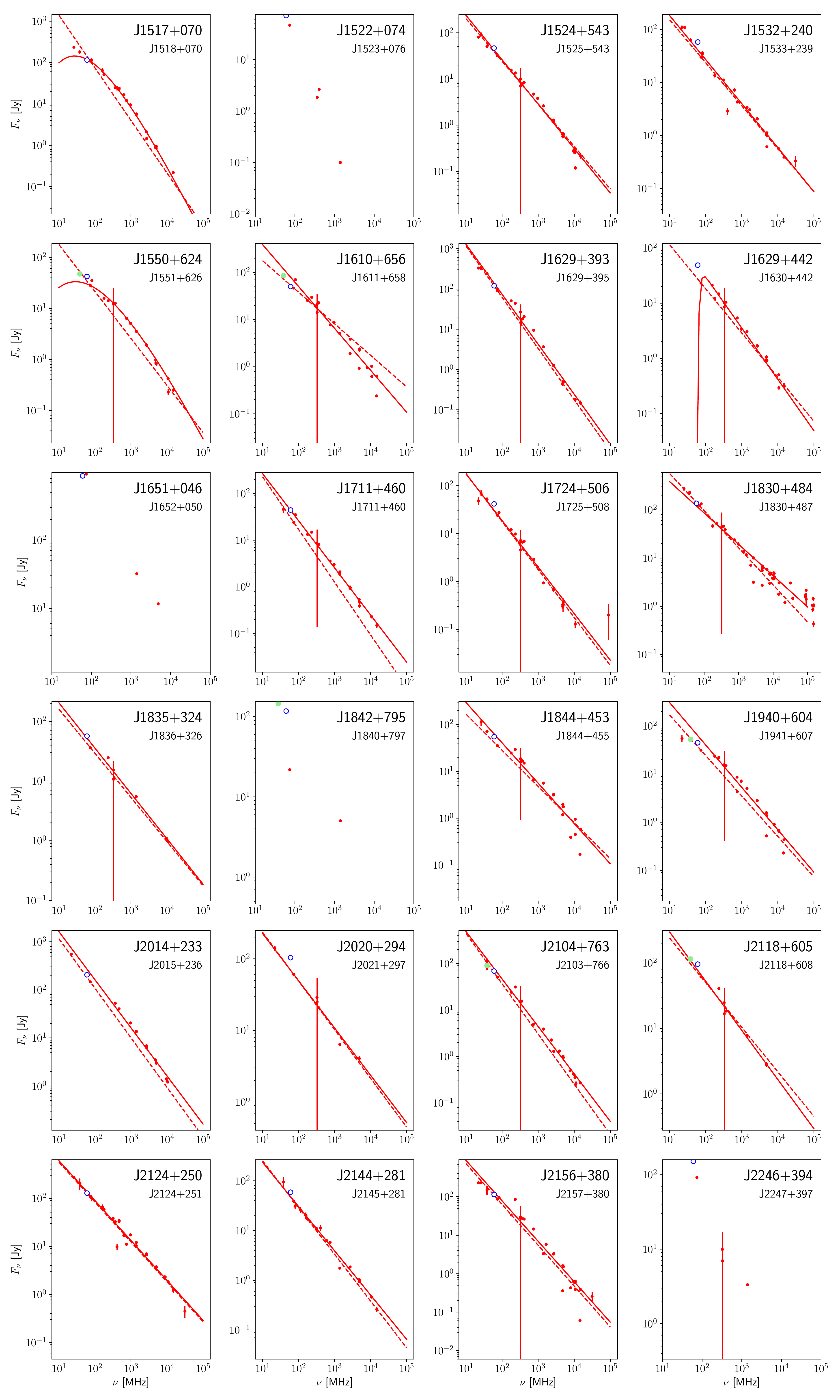}
\caption{Here the flux density as measured by AARTFAAC (open circles) are compared to the measurements and spectra fit by Helmbolt et al. (2008) (red dots and lines) measurements by Klein et al. (1979) (blue circles) and Rees et al. (1990) (green circles). The larger label refers to the designation assigned by Helmbolt et al. (2008), and the smaller label below is the label in the AARTFAAC catalogue.} 
\label{fig:helm_compare_104}
\end{center}
\end{figure}



\bsp	
\label{lastpage}
\end{document}